\documentclass[a4paper,11pt]{article}%
\usepackage{jheppub}
\usepackage{amsfonts}
\usepackage{amsmath}
\usepackage{amssymb}
\usepackage{graphicx}%
\providecommand{\U}[1]{\protect\rule{.1in}{.1in}}
\pdfoutput=1
\abstract{The NLO evolution equations for quadrupole and double dipole operators have been obtained within the high energy operator expansion method. The corresponding quasi-conformal evolution equations for the composite operators  were constructed.}
\arxivnumber{}
\affiliation{Budker Institute of Nuclear Physics and Novosibirsk State University,  630090 Novosibirsk, Russia}
\emailAdd{A.V.Grabovsky@inp.nsk.su}
\begin{document}

\title{\boldmath On the low-x NLO evolution of 4 point colorless operators}
\author{A. V. Grabovsky}
\maketitle

\flushbottom

\section{Introduction}

This paper develops the Wilson line approach to high energy scattering
proposed in \cite{Balitsky96} to the case of quadrupole and double dipole
operators in the next to leading order (NLO). Such operators naturally appear
when one studies amplitudes for diffractive processes with the production of 3
or 4 particles in the Regge limit. Moreover, the quadrupole operator enters
the definition of the Weizs\"{a}cker-Williams gluon distribution \cite{10},
\cite{11}, \cite{MueWW} which gives the Fock space number density of gluons
inside dense hadrons in light-cone gauge. One can find the NLO evolution
equation for the operator necessary for the Weizs\"{a}cker-Williams gluon
distribution differentiating the quadrupole equation obtained in this paper.
This result is going to be presented in a future work.

In the Wilson line approach to high energy scattering \cite{Balitsky96} the
amplitudes are convolutions of impact factors and a Green function. The impact
factors describe the decomposition of the colliding particles into quarks and
gluons while the Green function is responsible for the interaction of these
quarks and gluons with the quarks and gluons from the other colliding
particle. In this framework such fast-moving partons are depicted as Wilson
lines with the path going along their trajectories. Hence, the corresponding
Green functions are the operators constructed of the Wilson lines. These
operators obey the evolution equations with respect to the rapidity divide.
This rapidity divide separates the gluon field into the fast quantum one and
the slow external field of the other particle, through which the current quark
or gluon is propagating.

In the most thoroughly studied case of a virtual photon splitting into quark
antiquark pair, the corresponding Wilson line operator is a color dipole. The
evolution equation for this operator is known as the Balitsky - Kovchegov (BK)
equation \cite{Balitsky96}, \cite{kovchegov}. The NLO corrections to this
equation were calculated in \cite{Balitsky:2006wa}, \cite{koweigert},
\cite{Balitsky:2008zza}, \cite{Balitsky:2009xg}. Another interesting case is
application of this formalism to a proton. The proton has baryon color
structure and can be described as a 3-quark Wilson loop operator (3QWL). The
evolution equation for this operator was calculated in the leading order (LO)
in \cite{Gerasimov:2012bj} and in the NLO in \cite{bg}. The latter calculation
was based on the NLO hierarchy of the evolution equations for the Wilson lines
with open indices \cite{Balitsky:2013fea} and the connected contribution to
the 3QWL kernel \cite{Grabovsky:2013mba}. These results were also obtained in
the Jalilian-Marian, Iancu, McLerran, Weigert, Leonidov and Kovner (JIMWLK)
formalism \cite{jimwlk}. The hamiltonian equivalent to the NLO\ hierarchy was
obtained in \cite{Kovner:2013ona} and the evolution equation for the 3QWL in
\cite{Kovner:2014lca}. The NLO kernel for the evolution Wilson line operators
was also constructed in \cite{sch}.

The quadrupole and the double dipole are 4-particle colorless operators. Their
LO linear evolution equations were derived in \cite{MueWW},
\cite{Balitsky:2009xg}, \cite{2j}, \cite{9j}. Here the results of
\cite{Balitsky:2013fea} and \cite{Grabovsky:2013mba} are used to construct the
NLO evolution equations for these operators and the results of
\cite{Balitsky:2009xg} and \cite{bg} are used to check these equations.

The paper is organized as follows. The next section contains the definitions
and necessary results. Sections 3 and 4 present the NLO evolution equations
for the quadrupole and the double dipole operators in the standard and
quasi-conformal forms. Section 5 discusses different checks of the results.
Section 6 concludes the paper.

\section{Definitions and building blocks}

The light cone vectors $n_{1}$ and $n_{2}$ are defined as%
\begin{equation}
n_{1}=\left(  1,0,0,1\right)  ,\quad n_{2}=\frac{1}{2}\left(  1,0,0,-1\right)
,\quad n_{1}^{+}=n_{2}^{-}=n_{1}n_{2}=1
\end{equation}
and any vector $p$ can be decomposed as%
\begin{equation}
p^{+}=p_{-}=pn_{2}=\frac{1}{2}\left(  p^{0}+p^{3}\right)  ,\qquad p_{+}%
=p^{-}=pn_{1}=p^{0}-p^{3},
\end{equation}%
\begin{equation}
p=p^{+}n_{1}+p^{-}n_{2}+p_{\bot},\qquad p^{2}=2p^{+}p^{-}-\vec{p}^{\,2},
\end{equation}%
\begin{equation}
\quad p\,k=p^{\mu}k_{\mu}=p^{+}k^{-}+p^{-}k^{+}-\vec{p}\vec{k}=p_{+}%
k_{-}+p_{-}k_{+}-\vec{p}\vec{k}.
\end{equation}
For brevity the following notation for traces is used
\begin{equation}
tr(U_{i}U_{j}{}^{\dag}...U_{k}U_{l}{}^{\dag})\equiv\mathbf{U}_{ij^{\dag
}...kl^{\dag}}, \label{trace_Notation}%
\end{equation}
where%
\begin{equation}
U_{i}=U\left(  \vec{r}_{i},\eta\right)  =Pe^{ig\int_{-\infty}^{+\infty}%
b_{\eta}^{-}(r^{+},\vec{r})dr^{+}}, \label{WL}%
\end{equation}
and $b_{\eta}^{-}$ is the external shock wave field built from only slow
gluons
\begin{equation}
b_{\eta}^{-}=\int\frac{d^{4}p}{\left(  2\pi\right)  ^{4}}e^{-ipz}b^{-}\left(
p\right)  \theta(e^{\eta}-p^{+}).
\end{equation}
The parameter $\eta$ separates the slow gluons entering the Wilson lines from
the fast ones in the impact factors. The field%
\begin{equation}
b^{\mu}\left(  r\right)  =b^{-}(r^{+},\vec{r})n_{2}^{\mu}=\delta
(r^{+})b\left(  \vec{r}\right)  n_{2}^{\mu}.
\end{equation}
The coordinates $\vec{r}_{1,2,3,4}$ denote the quarks, and $\vec{r}_{0}%
,\vec{r}_{5}$ are the coordinates of the gluons. In intermediate formulas the
coordinates $\vec{r}_{6,7}$ will also be used. The $SU(N_{c})$ identities
\begin{equation}
U_{4}^{ba}=2tr(t^{b}U_{4}t^{a}U_{4}^{\dag}),\quad(t^{a})_{i}^{j}(t^{a}%
)_{k}^{l}=\frac{1}{2}\delta_{i}^{l}\delta_{k}^{j}-\frac{1}{2N_{c}}\delta
_{i}^{j}\delta_{k}^{l} \label{Uadjoint}%
\end{equation}
are necessary to rewrite the $SU(N_{c})$ operators only through the Wilson
lines in the fundamental representation. For a generic operator $O$ the
rapidity evolution equation has the form
\begin{equation}
\frac{\partial}{\partial\eta}\langle O\rangle=\langle K_{LO}\otimes
O\rangle+\langle K_{NLO}\otimes O\rangle.
\end{equation}
where $K_{LO}\sim\alpha_{s}$ and $K_{NLO}\sim\alpha_{s}^{2}.$ The
$\langle...\rangle$ brackets were explicitly written to denote that the
calculation was performed in the shockwave background. Hereafter they will be
often omitted to avoid overloading the notation. The BK equation in this
notation reads \cite{Balitsky96}%
\begin{equation}
\frac{\partial\mathbf{U}_{12^{\dag}}}{\partial\eta}={\frac{\alpha_{s}}%
{2\pi^{2}}}\!\int\!d\vec{r}_{0}\frac{\vec{r}_{12}{}^{2}}{\vec{r}_{10}{}%
^{2}\vec{r}_{20}{}^{2}}\left(  \mathbf{U}_{2^{\dag}0}\mathbf{U}_{0^{\dag}%
1}-N_{c}\mathbf{U}_{2^{\dag}1}\right)  , \label{LOdipole}%
\end{equation}
where $\vec{r}_{ij}=\vec{r}_{i}-\vec{r}_{j}.$ The LO quadrupole evolution
equation reads \cite{MueWW}%
\begin{align}
\frac{\partial\mathbf{U}_{12^{\dag}34^{\dag}}}{\partial\eta}=  &
{\frac{\alpha_{s}}{4\pi^{2}}}\!\int\!d\vec{r}_{0}\left\{  \frac{\vec{r}_{14}%
{}^{2}}{\vec{r}_{10}{}^{2}\vec{r}_{40}{}^{2}}\left(  \mathbf{U}_{10^{\dag}%
}\mathbf{U}_{02^{\dag}34^{\dag}}+\mathbf{U}_{4^{\dag}0}\mathbf{U}_{12^{\dag
}30^{\dag}}-\left(  0\rightarrow1\equiv0\rightarrow4\right)  \right)  \right.
\nonumber\\
+  &  \frac{\vec{r}_{12}{}^{2}}{\vec{r}_{10}{}^{2}\vec{r}_{20}{}^{2}}\left(
\mathbf{U}_{10^{\dag}}\mathbf{U}_{02^{\dag}34^{\dag}}+\mathbf{U}_{2^{\dag}%
0}\mathbf{U}_{10^{\dag}34^{\dag}}-\left(  0\rightarrow1\equiv0\rightarrow
2\right)  \right) \nonumber\\
-  &  \frac{\vec{r}_{24}{}^{2}}{2\vec{r}_{20}{}^{2}\vec{r}_{40}{}^{2}}\left(
\mathbf{U}_{10^{\dag}}\mathbf{U}_{02^{\dag}34^{\dag}}+\mathbf{U}_{30^{\dag}%
}\mathbf{U}_{04^{\dag}12^{\dag}}-(0\rightarrow4\equiv0\rightarrow2)\right)
\nonumber\\
-  &  \left.  \frac{\vec{r}_{13}{}^{2}}{2\vec{r}_{10}{}^{2}\vec{r}_{30}{}^{2}%
}\left(  \mathbf{U}_{4^{\dag}0}\mathbf{U}_{12^{\dag}30^{\dag}}+\mathbf{U}%
_{2^{\dag}0}\mathbf{U}_{34^{\dag}10^{\dag}}-(0\rightarrow1\equiv
0\rightarrow3)\right)  +(1\leftrightarrow3,2\leftrightarrow4)\right\}  .
\label{LOkernel}%
\end{align}
Here $\left(  0\rightarrow1\equiv0\rightarrow4\right)  $ stands for the
substitution $U_{0}\rightarrow U_{1}$ or $U_{0}\rightarrow U_{4},$ which gives
the same result. In addition $(1\leftrightarrow3,2\leftrightarrow4)$ means
that one has to change $\vec{r}_{1}\leftrightarrow\vec{r}_{3},\vec{r}%
_{2}\leftrightarrow\vec{r}_{4}$ and $U_{1}\leftrightarrow U_{3},U_{2}%
\leftrightarrow U_{4}.$ We will also need the LO evolution equations for the
double dipole, sextupole and the dipole-quadrupole product. All these
equations follow from the LO hierarchy \cite{Balitsky96} directly.
\begin{align}
\frac{\partial\mathbf{U}_{12^{\dag}}\mathbf{U}_{34^{\dag}}}{\partial\eta}=  &
{\frac{\alpha_{s}}{4\pi^{2}}}\!\int\!d\vec{r}_{0}\left(  \frac{\vec{r}_{13}%
{}^{2}}{\vec{r}_{10}{}^{2}\vec{r}_{30}{}^{2}}-\frac{\vec{r}_{23}{}^{2}}%
{\vec{r}_{20}{}^{2}\vec{r}_{30}{}^{2}}-\frac{\vec{r}_{14}{}^{2}}{\vec{r}%
_{10}{}^{2}\vec{r}_{40}{}^{2}}+\frac{\vec{r}_{24}{}^{2}}{\vec{r}_{20}{}%
^{2}\vec{r}_{40}{}^{2}}\right) \nonumber\\
\times &  (\mathbf{U}_{2^{\dag}14^{\dag}3}+\mathbf{U}_{2^{\dag}34^{\dag}%
1}-\mathbf{U}_{2^{\dag}10^{\dag}34^{\dag}0}-\mathbf{U}_{2^{\dag}04^{\dag
}30^{\dag}1})+\mathbf{U}_{4^{\dag}3}\frac{\partial\mathbf{U}_{12^{\dag}}%
}{\partial\eta}+\mathbf{U}_{2^{\dag}1}\frac{\partial\mathbf{U}_{4^{\dag}3}%
}{\partial\eta}. \label{LOdoubledipole}%
\end{align}%
\begin{align}
\frac{\partial\mathbf{U}_{12^{\dag}34^{\dag}}\mathbf{U}_{76^{\dag}}}%
{\partial\eta}=  &  {\frac{\alpha_{s}}{4\pi^{2}}}\!\int\!d\vec{r}_{0}\left\{
\left(  \mathbf{U}_{0^{\dag}76^{\dag}02^{\dag}34^{\dag}1}+\mathbf{U}_{0^{\dag
}12^{\dag}34^{\dag}06^{\dag}7}-\left(  0\rightarrow7\equiv0\rightarrow
6\right)  \right)  \right. \nonumber\\
\times &  \left(  \frac{\vec{r}_{16}{}^{2}}{\vec{r}_{01}{}^{2}\vec{r}_{06}%
{}^{2}}-\frac{\vec{r}_{17}{}^{2}}{\vec{r}_{01}{}^{2}\vec{r}_{07}{}^{2}%
}\right)  +\left(  \frac{\vec{r}_{27}{}^{2}}{\vec{r}_{02}{}^{2}\vec{r}_{07}%
{}^{2}}-\frac{\vec{r}_{26}{}^{2}}{\vec{r}_{02}{}^{2}\vec{r}_{06}{}^{2}}\right)
\nonumber\\
\times &  \left(  \mathbf{U}_{0^{\dag}76^{\dag}02^{\dag}34^{\dag}1}%
+\mathbf{U}_{0^{\dag}34^{\dag}12^{\dag}06^{\dag}7}-\left(  0\rightarrow
7\equiv0\rightarrow6\right)  \right) \nonumber\\
+  &  (1\leftrightarrow3,2\leftrightarrow4)\}+\mathbf{U}_{76^{\dag}}%
\frac{\partial\mathbf{U}_{12^{\dag}34^{\dag}}}{\partial\eta}+\mathbf{U}%
_{12^{\dag}34^{\dag}}\frac{\partial\mathbf{U}_{76^{\dag}}}{\partial\eta}.
\label{LOdipquad}%
\end{align}%
\begin{align}
\frac{\partial\mathbf{U}_{12^{\dag}34^{\dag}56^{\dag}}}{\partial\eta}=  &
{\frac{\alpha_{s}}{4\pi^{2}}}\!\int\!d\vec{r}_{0}\{\frac{\vec{r}_{25}{}^{2}%
}{\vec{r}_{02}{}^{2}\vec{r}_{05}{}^{2}}(\mathbf{U}_{0^{\dag}34^{\dag}%
5}\mathbf{U}_{2^{\dag}06^{\dag}1}+\mathbf{U}_{0^{\dag}56^{\dag}1}%
\mathbf{U}_{2^{\dag}34^{\dag}0}-\left(  0\rightarrow5\equiv0\rightarrow
2\right)  )\nonumber\\
-  &  \frac{\vec{r}_{15}{}^{2}}{\vec{r}_{01}{}^{2}\vec{r}_{05}{}^{2}%
}(\mathbf{U}_{0^{\dag}56^{\dag}1}\mathbf{U}_{2^{\dag}34^{\dag}0}%
+\mathbf{U}_{6^{\dag}0}\mathbf{U}_{0^{\dag}12^{\dag}34^{\dag}5}-\left(
0\rightarrow5\equiv0\rightarrow1\right)  )\nonumber\\
-  &  \frac{\vec{r}_{26}{}^{2}}{\vec{r}_{02}{}^{2}\vec{r}_{06}{}^{2}%
}(\mathbf{U}_{0^{\dag}34^{\dag}5}\mathbf{U}_{2^{\dag}06^{\dag}1}%
+\mathbf{U}_{0^{\dag}1}\mathbf{U}_{2^{\dag}34^{\dag}56^{\dag}0}-\left(
0\rightarrow2\equiv0\rightarrow6\right)  )\nonumber\\
+  &  \frac{\vec{r}_{16}{}^{2}}{\vec{r}_{01}{}^{2}\vec{r}_{06}{}^{2}%
}(\mathbf{U}_{6^{\dag}0}\mathbf{U}_{0^{\dag}12^{\dag}34^{\dag}5}%
+\mathbf{U}_{0^{\dag}1}\mathbf{U}_{2^{\dag}34^{\dag}56^{\dag}0}-\left(
0\rightarrow1\equiv0\rightarrow6\right)  )\nonumber\\
+  &  \frac{\vec{r}_{12}{}^{2}}{\vec{r}_{01}{}^{2}\vec{r}_{02}{}^{2}%
}(\mathbf{U}_{2^{\dag}0}\mathbf{U}_{0^{\dag}34^{\dag}56^{\dag}1}%
+\mathbf{U}_{0^{\dag}1}\mathbf{U}_{2^{\dag}34^{\dag}56^{\dag}0}-\left(
0\rightarrow1\equiv0\rightarrow2\right)  )\nonumber\\
+  &  (1\rightarrow3\rightarrow5\rightarrow1,2\rightarrow4\rightarrow
6\rightarrow2)+(1\rightarrow5\rightarrow3\rightarrow1,2\rightarrow
6\rightarrow4\rightarrow2)\}. \label{sex}%
\end{align}
Here $1\rightarrow3\rightarrow5\rightarrow1$ stands for permutation, i.e. one
has to change $\vec{r}_{1}\rightarrow\vec{r}_{3},$ $\vec{r}_{3}\rightarrow
\vec{r}_{5},$ $\vec{r}_{5}\rightarrow\vec{r}_{1}$ and $U_{1}\rightarrow
U_{3},$ $U_{3}\rightarrow U_{5},$ $U_{5}\rightarrow U_{1}.$

For the self and the pairwise NLO interactions one can take the results of
\cite{Balitsky:2013fea} while the triple-interaction diagrams were already
calculated in \cite{Grabovsky:2013mba}. The results of these papers were
derived using sharp cutoff on the rapidity variable. Since this paper is
devoted to color singlet operators one can drop the kernels which vanish
acting on the colorless operators, as was shown in \cite{Kovner:2013ona}. The
rest reads%
\begin{align}
\left.  \frac{\partial\left(  U_{1}\right)  _{i}^{j}}{\partial\eta}\right\vert
_{NLO}  &  =\frac{\alpha_{s}^{2}}{8\pi^{4}}\int d\vec{r}_{5}d\vec{r}%
_{0}J_{11105}[if^{ad^{\prime}e^{\prime}}(\{t^{d}t^{e}\}U_{1}t^{a})_{i}%
^{j}-if^{ade}(t^{a}U_{1}\{t^{d^{\prime}}t^{e^{\prime}}\})_{i}^{j}]\nonumber\\
&  \times U_{0}^{dd^{\prime}}(U_{5}^{ee^{\prime}}-U_{0}^{ee^{\prime}}%
)+\frac{\alpha_{s}^{2}N_{c}}{4\pi^{3}}\int\frac{d\vec{r}_{5}}{\vec{r}_{15}%
{}^{4}}(U_{5}^{ab}-U_{1}^{ab})(t^{a}U_{1}t^{b})_{i}^{j}\beta\ln\left(
\frac{\vec{r}_{15}^{\,\,2}}{\tilde{\mu}^{2}}\right)  , \label{Uinitial}%
\end{align}%
\begin{equation}
\beta=\left(  \frac{11}{3}-\frac{2}{3}\frac{n_{f}}{N_{c}}\right)  ,\quad
\beta\ln\frac{1}{\tilde{\mu}^{2}}=\left(  \frac{11}{3}-\frac{2}{3}\frac{n_{f}%
}{N_{c}}\right)  \ln\left(  \frac{\mu^{2}}{4e^{2\psi\left(  1\right)  }%
}\right)  +\frac{67}{9}-\frac{\pi^{2}}{3}-\frac{10}{9}\frac{n_{f}}{N_{c}},
\end{equation}
$n_{f}$ is the number of the quark flavours, $\mu^{2}$ is the renormalization
scale in the $\overline{MS}$-scheme and $J_{ijklm}\equiv J(\vec{r}_{i},\vec
{r}_{j},\vec{r}_{k},\vec{r}_{l},\vec{r}_{m})$%
\begin{align}
J_{12305}=  &  \left(  \frac{\vec{r}_{01}\vec{r}_{52}}{\vec{r}_{01}{}^{2}%
\vec{r}_{05}{}^{2}\vec{r}_{25}{}^{2}}+\frac{2\left(  \vec{r}_{01}\vec{r}%
_{03}\right)  \left(  \vec{r}_{05}\vec{r}_{25}\right)  }{\vec{r}_{01}{}%
^{2}\vec{r}_{03}{}^{2}\vec{r}_{05}{}^{2}\vec{r}_{25}{}^{2}}-\frac{2\left(
\vec{r}_{01}\vec{r}_{03}\right)  \left(  \vec{r}_{25}\vec{r}_{35}\right)
}{\vec{r}_{01}{}^{2}\vec{r}_{03}{}^{2}\vec{r}_{25}{}^{2}\vec{r}_{35}{}^{2}%
}+\frac{2\left(  \vec{r}_{01}\vec{r}_{05}\right)  \left(  \vec{r}_{25}\vec
{r}_{35}\right)  }{\vec{r}_{01}{}^{2}\vec{r}_{05}{}^{2}\vec{r}_{25}{}^{2}%
\vec{r}_{35}{}^{2}}\right) \nonumber\\
\times &  \ln\left(  \frac{\vec{r}_{03}{}^{2}}{\vec{r}_{35}{}^{2}}\right)  .
\label{J}%
\end{align}
This function has the properties%
\begin{equation}
J_{ijk05}=-J_{jik50},\quad J_{11105}=\frac{\left(  \vec{r}_{51}\vec{r}%
_{01}\right)  }{\vec{r}_{01}{}^{2}\vec{r}_{05}{}^{2}\vec{r}_{15}{}^{2}}%
\ln\left(  \frac{\vec{r}_{01}{}^{2}}{\vec{r}_{15}{}^{2}}\right)  .
\end{equation}%
\begin{equation}
\left.  \frac{\partial\left(  U_{1}\right)  _{i}^{j}\left(  U_{2}\right)
_{k}^{l}}{\partial\eta}\right\vert _{NLO}=\frac{\alpha_{s}^{2}}{8\pi^{4}}\int
d\vec{r}_{5}d\vec{r}_{0}\left(  \mathcal{A}_{1}+\mathcal{A}_{2}+\mathcal{A}%
_{3}\right)  +\frac{\alpha_{s}^{2}}{8\pi^{3}}\int d\vec{r}_{5}\left(
\mathcal{B}_{1}+N_{c}\mathcal{B}_{2}\right)  . \label{UUinitial}%
\end{equation}
Here%
\begin{align}
\mathcal{A}_{1}=  &  [(t^{a}U_{1})_{i}^{j}(U_{2}t^{b})_{k}^{l}+(t^{a}%
U_{2})_{k}^{l}(U_{1}t^{b})_{i}^{j}]\left[  f^{ade}f^{bd^{\prime}e^{\prime}%
}U_{0}^{dd^{\prime}}(U_{5}^{ee^{\prime}}-U_{0}^{ee^{\prime}})4L_{12}\right.
\nonumber\\
+  &  \left.  4n_{f}L_{12}^{q}tr(t^{a}U_{5}t^{b}(U_{0}^{\dag}-U_{5}^{\dag
}))\right]  , \label{A1}%
\end{align}
where $L_{ij}\equiv L(\vec{r}_{i},\vec{r}_{j})$ and $L_{ij}^{q}\equiv
L^{q}(\vec{r}_{i},\vec{r}_{j})$ were introduced in this form in \cite{bg}
\begin{align}
L_{12}  &  =\left[  \frac{1}{\vec{r}_{01}{}^{2}\vec{r}_{25}{}^{2}-\vec{r}%
_{02}{}^{2}\vec{r}_{15}{}^{2}}\left(  -\frac{\vec{r}_{12}{}^{4}}{8}\left(
\frac{1}{\vec{r}_{01}{}^{2}\vec{r}_{25}{}^{2}}+\frac{1}{\vec{r}_{02}{}^{2}%
\vec{r}_{15}{}^{2}}\right)  +\frac{\vec{r}_{12}{}^{2}}{\vec{r}_{05}{}^{2}%
}-\frac{\vec{r}_{02}{}^{2}\vec{r}_{15}{}^{2}+\vec{r}_{01}{}^{2}\vec{r}_{25}%
{}^{2}}{4\vec{r}_{05}{}^{4}{}}\right)  \right. \nonumber\\
&  +\left.  \frac{\vec{r}_{12}{}^{2}}{8\vec{r}_{05}{}^{2}}\left(  \frac
{1}{\vec{r}_{02}{}^{2}\vec{r}_{15}{}^{2}}-\frac{1}{\vec{r}_{01}{}^{2}\vec
{r}_{25}{}^{2}}\right)  \right]  \ln\left(  \frac{\vec{r}_{01}{}^{2}\vec
{r}_{25}{}^{2}}{\vec{r}_{15}{}^{2}\vec{r}_{02}{}^{2}}\right)  +\frac{1}%
{2\vec{r}_{05}{}^{4}{}},\label{L12}\\
L_{12}^{q}  &  =\frac{1}{\vec{r}_{05}{}^{4}}\left\{  \frac{\vec{r}_{02}{}%
^{2}\vec{r}_{15}{}^{2}+\vec{r}_{01}{}^{2}\vec{r}_{25}{}^{2}-\vec{r}_{05}{}%
^{2}\vec{r}_{12}{}^{2}}{2(\vec{r}_{02}{}^{2}\vec{r}_{15}{}^{2}-\vec{r}_{01}%
{}^{2}\vec{r}_{25}{}^{2})}\ln\left(  \frac{\vec{r}_{02}{}^{2}\vec{r}_{15}%
{}^{2}}{\vec{r}_{01}{}^{2}\vec{r}_{25}{}^{2}}\right)  -1\right\}  .
\label{L12q}%
\end{align}
These functions have the unintegrable singularity at $\vec{r}_{05}=0$, which
is canceled by the subtraction in the color structure. They are symmetric
conformally invariant functions $L_{ij}^{(q)}=L_{ji}^{(q)}=L_{ij}^{(q)}%
|_{\vec{r}_{0}\leftrightarrow\vec{r}_{5}}$.%
\begin{align}
\mathcal{A}_{2}=  &  4(U_{0}-U_{1})^{dd^{\prime}}(U_{5}-U_{2})^{ee^{\prime}%
}\left\{  i[f^{ad^{\prime}e^{\prime}}(t^{d}U_{1}t^{a})_{i}^{j}(t^{e}U_{2}%
)_{k}^{l}-f^{ade}(t^{a}U_{1}t^{d^{\prime}})_{i}^{j}(U_{2}t^{e^{\prime}}%
)_{k}^{l}]J_{12105}\right. \nonumber\\
+  &  \left.  i[f^{ad^{\prime}e^{\prime}}(t^{d}U_{1})_{i}^{j}(t^{e}U_{2}%
t^{a})_{k}^{l}-f^{ade}(U_{1}t^{d^{\prime}})_{i}^{j}(t^{a}U_{2}t^{e^{\prime}%
})_{k}^{l}]J_{12205}\right\}  ,\label{A2}\\
\mathcal{A}_{3}=  &  2U_{0}^{dd^{\prime}}\left\{  i[f^{ad^{\prime}e^{\prime}%
}(U_{1}t^{a})_{i}^{j}(t^{d}t^{e}U_{2})_{k}^{l}-f^{ade}(t^{a}U_{1})_{i}%
^{j}(U_{2}t^{e^{\prime}}t^{d^{\prime}})_{k}^{l}]\right.  (U_{5}-U_{2}%
)^{ee^{\prime}}\nonumber\\
\times &  (J_{22105}+J_{21205}-J_{12205})+(J_{12105}+J_{11205}-J_{21105}%
)\nonumber\\
\times &  \left.  i[f^{ad^{\prime}e^{\prime}}(t^{d}t^{e}U_{1})_{i}^{j}%
(U_{2}t^{a})_{k}^{l}-f^{ade}(U_{1}t^{e^{\prime}}t^{d^{\prime}})_{i}^{j}%
(t^{a}U_{2})_{k}^{l}](U_{5}-U_{1})^{ee^{\prime}}\right\}  . \label{A3}%
\end{align}%
\begin{align}
\mathcal{B}_{1}=  &  2\ln\left(  \frac{\vec{r}_{15}^{\,\,2}}{\vec{r}%
_{12}^{\,\,2}}\right)  \ln\left(  \frac{\vec{r}_{25}^{\,\,2}}{\vec{r}%
_{12}^{\,\,2}}\right) \nonumber\\
\times &  \left\{  (U_{5}-U_{1})^{ab}i[f^{bde}(t^{a}U_{1}t^{d})_{i}^{j}%
(U_{2}t^{e})_{k}^{l}+f^{ade}(t^{e}U_{1}t^{b})_{i}^{j}(t^{d}U_{2})_{k}%
^{l}]\left(  \frac{(\vec{r}_{15}\vec{r}_{25})}{\vec{r}_{15}^{\,\,2}\vec
{r}_{25}^{\,\,2}}-\frac{1}{\vec{r}_{15}^{\,\,2}}\right)  \right. \nonumber\\
+  &  \left.  (U_{5}-U_{2})^{ab}i[f^{bde}(U_{1}t^{e})_{i}^{j}(t^{a}U_{2}%
t^{d})_{k}^{l}+f^{ade}(t^{d}U_{1})_{i}^{j}(t^{e}U_{2}t^{b})_{k}^{l}]\left(
\frac{(\vec{r}_{15}\vec{r}_{25})}{\vec{r}_{15}^{\,\,2}\vec{r}_{25}^{\,\,2}%
}-\frac{1}{\vec{r}_{25}^{\,\,2}}\right)  \right\}  ,\label{B1}\\
\mathcal{B}_{2}=  &  \beta\left(  2U_{5}-U_{1}-U_{2}\right)  ^{ab}[(t^{a}%
U_{1})_{i}^{j}(U_{2}t^{b})_{k}^{l}+(U_{1}t^{b})_{i}^{j}(t^{a}U_{2})_{k}%
^{l}]\nonumber\\
\times &  \left\{  \frac{(\vec{r}_{15}\vec{r}_{25})}{\vec{r}_{15}^{\,\,2}%
\vec{r}_{25}^{\,\,2}}\ln\left(  \frac{\vec{r}_{12}^{\,\,2}}{\tilde{\mu}^{2}%
}\right)  +\frac{1}{2\vec{r}_{15}^{\,\,2}}\ln\left(  \frac{\vec{r}%
_{25}^{\,\,2}}{\vec{r}_{12}^{\,\,2}}\right)  +\frac{1}{2\vec{r}_{25}^{\,\,2}%
}\ln\left(  \frac{\vec{r}_{15}^{\,\,2}}{\vec{r}_{12}^{\,\,2}}\right)
\right\}  . \label{B2}%
\end{align}%
\begin{align}
\left.  \frac{\partial}{\partial\eta}\left(  U_{1}\right)  _{i}^{j}\left(
U_{2}\right)  _{k}^{l}\left(  U_{3}\right)  _{m}^{n}\right\vert _{NLO}  &
=\frac{i\alpha_{s}^{2}}{2\pi^{4}}\int d\vec{r}_{5}d\vec{r}_{0}\nonumber\\
&  \times\{f^{cde}[\left(  t^{a}U_{1}\right)  _{i}^{j}(t^{b}U_{2})_{k}%
^{l}\left(  U_{3}t^{c}\right)  _{m}^{n}(U_{0}-U_{1})^{ad}(U_{5}-U_{2}%
)^{be}\nonumber\\
&  -\left(  U_{1}t^{a}\right)  _{i}^{j}(U_{2}t^{b})_{k}^{l}\left(  t^{c}%
U_{3}\right)  _{m}^{n}(U_{0}-U_{1})^{da}(U_{5}-U_{2})^{eb}]J_{12305}%
\nonumber\\
&  +f^{ade}[\left(  U_{1}t^{a}\right)  _{i}^{j}(t^{b}U_{2})_{k}^{l}\left(
t^{c}U_{3}\right)  _{m}^{n}(U_{0}-U_{3})^{cd}(U_{5}-U_{2})^{be}\nonumber\\
&  -\left(  t^{a}U_{1}\right)  _{i}^{j}(U_{2}t^{b})_{k}^{l}\left(  U_{3}%
t^{c}\right)  _{m}^{n}(U_{0}-U_{3})^{dc}(U_{5}-U_{2})^{eb}]J_{32105}%
\nonumber\\
&  +f^{bde}[\left(  t^{a}U_{1}\right)  _{i}^{j}(U_{2}t^{b})_{k}^{l}\left(
t^{c}U_{3}\right)  _{m}^{n}(U_{0}-U_{1})^{ad}(U_{5}-U_{3})^{ce}\nonumber\\
&  -\left(  U_{1}t^{a}\right)  _{i}^{j}(t^{b}U_{2})_{k}^{l}\left(  U_{3}%
t^{c}\right)  _{m}^{n}(U_{0}-U_{1})^{da}(U_{5}-U_{3})^{ec}]J_{13205}\}.
\label{UUUinitial}%
\end{align}
We will also need the following functions. The function $M_{i}^{jk}\equiv
M(\vec{r}_{i},\vec{r}_{j},\vec{r}_{k})$ was introduced in \cite{bg}%
\begin{align}
M_{2}^{13}  &  =\frac{1}{2}(J_{12205}+J_{23205}-J_{13205}-J_{22205}%
)\nonumber\\
&  =\frac{1}{4\vec{r}_{01}{}^{2}\vec{r}_{35}{}^{2}}\left(  \frac{\vec{r}%
_{12}{}^{2}\vec{r}_{23}{}^{2}}{\vec{r}_{02}{}^{2}\vec{r}_{25}{}^{2}}%
-\frac{\vec{r}_{15}{}^{2}\vec{r}_{23}{}^{2}}{\vec{r}_{05}{}^{2}\vec{r}_{25}%
{}^{2}}-\frac{\vec{r}_{03}{}^{2}\vec{r}_{12}{}^{2}}{\vec{r}_{02}{}^{2}\vec
{r}_{05}{}^{2}}+\frac{\vec{r}_{13}{}^{2}}{\vec{r}_{05}{}^{2}}\right)
\ln\left(  \frac{\vec{r}_{02}{}^{2}}{\vec{r}_{25}{}^{2}}\right)  . \label{M}%
\end{align}
It was also introduced as $M_{2}$ in \cite{app}. It has the property%
\begin{equation}
M_{k}^{ij}|_{5\leftrightarrow0}=-M_{k}^{ji}. \label{Masimproperty}%
\end{equation}
The functions $\tilde{L}_{ij}\equiv\tilde{L}(\vec{r}_{i},\vec{r}_{j})$ and
$M_{ij}\equiv M(\vec{r}_{i},\vec{r}_{j})$ were introduced in \cite{bg} as well%
\begin{equation}
\tilde{L}_{12}=\frac{1}{2}(M_{1}^{22}-M_{2}^{11})=\frac{\vec{r}_{12}{}^{2}%
}{8\vec{r}_{05}{}^{2}}\left[  \frac{\vec{r}_{12}{}^{2}\vec{r}_{05}{}^{2}}%
{\vec{r}_{01}{}^{2}\vec{r}_{02}{}{}^{2}\vec{r}_{15}{}^{2}\vec{r}_{25}{}^{2}%
}-\frac{1}{\vec{r}_{01}{}^{2}\vec{r}_{25}{}^{2}}-\frac{1}{\vec{r}_{02}{}%
^{2}\vec{r}_{15}{}^{2}}\right]  \ln\left(  \frac{\vec{r}_{01}{}^{2}\vec
{r}_{25}{}^{2}}{\vec{r}_{15}{}^{2}\vec{r}_{02}{}^{2}}\right)  , \label{Ltilde}%
\end{equation}%
\begin{equation}
M_{12}=\frac{1}{4}(M_{2}^{11}+M_{1}^{22})=\frac{\vec{r}_{12}{}^{2}}{16\vec
{r}_{05}{}^{2}}\left[  \frac{\vec{r}_{12}{}^{2}\vec{r}_{05}{}^{2}}{\vec
{r}_{01}{}^{2}\vec{r}_{02}{}{}^{2}\vec{r}_{15}{}^{2}\vec{r}_{25}{}^{2}}%
-\frac{1}{\vec{r}_{01}{}^{2}\vec{r}_{25}{}^{2}}-\frac{1}{\vec{r}_{02}{}%
^{2}\vec{r}_{15}{}^{2}}\right]  \ln\left(  \frac{\vec{r}_{01}{}^{2}\vec
{r}_{02}{}^{2}}{\vec{r}_{15}{}^{2}\vec{r}_{25}{}^{2}}\right)  . \label{Mij}%
\end{equation}
Here $\tilde{L}_{ij}$ is conformally invariant. Moreover, $\tilde{L}_{ij}$ is
antisymmetric w.r.t. both $5\leftrightarrow0$ and $i\leftrightarrow j$
transformations while $M_{ij}$ is antisymmetric w.r.t. $5\leftrightarrow0$.
One can also combine all the terms $\sim\beta$ into $M_{ij}^{\beta}\equiv
M^{\beta}(\vec{r}_{i},\vec{r}_{j})$%
\begin{equation}
M_{12}^{\beta}=\frac{N_{c}\beta}{2}\left\{  \ln\left(  \frac{\vec{r}%
_{12}^{\,\,2}}{\tilde{\mu}^{2}}\right)  +\frac{\vec{r}_{01}^{\,\,2}\vec
{r}_{02}^{\,\,2}}{\vec{r}_{12}^{\,\,2}}\ln\left(  \frac{\vec{r}_{02}^{\,\,2}%
}{\vec{r}_{01}^{\,\,2}}\right)  \left(  \frac{1}{\vec{r}_{02}^{\,\,2}}%
-\frac{1}{\vec{r}_{01}^{\,\,2}}\right)  \right\}  . \label{Mbeta}%
\end{equation}
The NLO BK kernel reads \cite{Balitsky:2009xg}%
\begin{align}
\langle K_{NLO}\otimes\mathbf{U}_{12^{\dag}}\rangle=  &  {\frac{\alpha_{s}%
^{2}}{4\pi^{3}}}\!\int\!d\vec{r}_{0}\frac{\vec{r}_{12}{}^{2}}{\vec{r}_{10}%
{}^{2}\vec{r}_{20}{}^{2}}\left\{  M_{12}^{\beta}-N_{c}\ln\left(  \frac{\vec
{r}_{12}{}^{2}}{\vec{r}_{10}{}^{2}}\right)  \ln\left(  \frac{\vec{r}_{12}%
{}^{2}}{\vec{r}_{20}{}^{2}}\right)  \right\} \nonumber\\
\times &  \left(  \mathbf{U}_{2^{\dag}0}\mathbf{U}_{0^{\dag}1}-N_{c}%
\mathbf{U}_{2^{\dag}1}\right)  +{\frac{\alpha_{s}^{2}}{4\pi^{4}}}\!\int
\!d\vec{r}_{0}d\vec{r}_{5}\{\tilde{L}_{12}(\mathbf{U}_{0^{\dag}5}%
\mathbf{U}_{2^{\dag}0}\mathbf{U}_{5^{\dag}1}-(0\leftrightarrow5))\nonumber\\
+  &  L_{12}((\mathbf{U}_{0^{\dag}52^{\dag}05^{\dag}1}-\mathbf{U}_{0^{\dag}%
1}\mathbf{U}_{2^{\dag}5}\mathbf{U}_{5^{\dag}0}-(0\rightarrow
5))+(0\leftrightarrow5))\nonumber\\
-  &  2n_{f}L_{12}^{q}(tr(t^{a}U_{1}t^{b}U_{2}^{\dag})tr(t^{a}U_{5}t^{b}%
(U_{0}^{\dag}-U_{5}^{\dag}))+(5\leftrightarrow0))\}. \label{NLO_BK}%
\end{align}

\section{Construction of the kernel}

First, one has to discuss the singularities of the building blocks from the
previous section. All the ultraviolet (UV) singularities in (\ref{Uinitial}),
(\ref{UUinitial}), and (\ref{UUUinitial}) were removed by the renormalization.
It means that these expressions converge at $\vec{r}_{0}=\vec{r}_{1,2,3,4,5}$
and $\vec{r}_{5}=\vec{r}_{0,1,2,3,4}.$ In particular, the functions $J$ in
$\mathcal{A}_{2}$ (\ref{A2}), $\mathcal{A}_{3}$ (\ref{A3}), and
(\ref{UUUinitial}) are convergent at these points, which ensures UV-safety of
these expressions. However, the function $J_{11105}$ in the first line of
(\ref{Uinitial}), has the UV singularity at $\vec{r}_{0}=\vec{r}_{5}=\vec
{r}_{1}.$ As in (\ref{L12})\ this singularity is removed by the subtraction in
the color structure.

Nevertheless, these expressions have infrared (IR) singularities, which appear
as both $\vec{r}_{0,5}\rightarrow\infty.$ Indeed, changing the variables e.g.
as $r_{0}=ut,$ $r_{5}=u\bar{t},\bar{t}=1-t,$ one faces a logarithmic
singularity integrating w.r.t. $u$ first%
\begin{equation}
\int d\vec{r}_{5}d\vec{r}_{0}J_{12305}=\int d\phi_{5}d\phi_{0}\int_{0}%
^{1}dt\int_{0}^{+\infty}du\left(  \frac{2\cos(\phi_{05})}{u(t^{2}+\bar{t}%
^{2}-2t\bar{t}\cos(\phi_{05}))}\ln\left(  \frac{t}{\bar{t}}\right)  +O\left(
\frac{1}{u^{2}}\right)  \right)  .
\end{equation}
Hence this double integral is ill-defined and requires either regularization
or definition in terms of the iterated integrals. To understand how to
correctly treat the IR singularities one can either return to the diagrams and
keep the regularization, or calculate the known dipole equation and fix the
definition from there. The latter way is attempted here. Assembling BK kernel
(\ref{NLO_BK}) from (\ref{Uinitial}--\ref{UUinitial}), one can see that all
the $\beta$-functional terms go to $M_{12}^{\beta}$, $\mathcal{A}_{1}$
(\ref{A1}) reshapes to the terms $\sim L_{12},L_{12}^{q},$ the Wilson line
operators from (\ref{Uinitial}), (\ref{A2}--\ref{A3}) depending on both
$\vec{r}_{5}$ and $\vec{r}_{0}$ give the term $\sim\tilde{L}_{12}\ $after the
symmetrization%
\begin{align}
A(U)F(\vec{r})  &  \rightarrow\left[  AF\right]  ^{sym}\nonumber\\
&  =\frac{\left[  A+A\left(  0\leftrightarrow5\right)  \right]  \left[
F+F\left(  0\leftrightarrow5\right)  \right]  +\left[  A-A\left(
0\leftrightarrow5\right)  \right]  \left[  F-F\left(  0\leftrightarrow
5\right)  \right]  }{4}. \label{step1}%
\end{align}
Next, $\mathcal{B}_{1}$ (\ref{B1}) gives one half of the double logarithm
contribution. All the remaining terms are to be equal to the other half of the
double logarithm contribution. They read
\begin{align}
\Delta K=  &  {\frac{\alpha_{s}^{2}N_{c}}{16\pi^{4}}}\!\int\!d\vec{r}_{0}%
d\vec{r}_{5}\{\left(  \mathbf{U}_{2^{\dag}5}\mathbf{U}_{5^{\dag}1}%
-\mathbf{U}_{0^{\dag}1}\mathbf{U}_{2^{\dag}0}\right)  \left(  J_{22105}%
+J_{11205}\right) \nonumber\\
+  &  \left(  J_{21105}-J_{12105}+J_{12205}-J_{21205}\right)  \left(
2N_{c}\mathbf{U}_{2^{\dag}1}-\mathbf{U}_{0^{\dag}1}\mathbf{U}_{2^{\dag}%
0}-\mathbf{U}_{2^{\dag}5}\mathbf{U}_{5^{\dag}1}\right)  \}. \label{deltaK}%
\end{align}
This term is IR safe. The second line is the product of 2 expressions
symmetric w.r.t. $0\leftrightarrow5$ permutation. Therefore one can set
$U_{5}\rightarrow U_{0}$ there. In the first line there is a product of 2
expressions antisymmetric w.r.t. $0\leftrightarrow5$ permutation. Hence, one
could add and subtract $N_{c}\mathbf{U}_{2^{\dag}1}$ in the first brackets and
write
\begin{align}
\Delta K\rightarrow &  {\frac{\alpha_{s}^{2}N_{c}}{8\pi^{4}}}\!\int\!d\vec
{r}_{0}d\vec{r}_{5}\{\left(  N_{c}\mathbf{U}_{2^{\dag}1}-\mathbf{U}_{2^{\dag
}0}\mathbf{U}_{0^{\dag}1}\right)  \left(  J_{22105}+J_{11205}\right)
\nonumber\\
+  &  \left(  J_{21105}+J_{12205}-J_{12105}-J_{21205}\right)  \left(
N_{c}\mathbf{U}_{2^{\dag}1}-\mathbf{U}_{0^{\dag}1}\mathbf{U}_{2^{\dag}%
0}\right)  \}\\
=  &  {\frac{\alpha_{s}^{2}N_{c}}{8\pi^{4}}}\!\int\!d\vec{r}_{0}\int d\vec
{r}_{5}\left(  N_{c}\mathbf{U}_{2^{\dag}1}-\mathbf{U}_{0^{\dag}1}%
\mathbf{U}_{2^{\dag}0}\right) \nonumber\\
\times &  \{J_{22105}-J_{21205}+J_{11205}-J_{12105}+J_{21105}+J_{12205}).
\end{align}
One could understand the latter integral as an iterated one. Then, using the
integrals%
\begin{equation}
\int\frac{d\vec{r}_{5}}{\pi}(J_{ijk05}-J_{ikj05})=\left(  \frac{\vec{r}%
_{0i}\vec{r}_{0j}}{\vec{r}_{0i}^{\,\,2}\vec{r}_{0j}^{\,\,2}}-\frac{\vec
{r}_{0i}\vec{r}_{0k}}{\vec{r}_{0i}^{\,\,2}\vec{r}_{0k}^{\,\,2}}\right)
\ln\left(  \frac{\vec{r}_{jk}^{\,\,2}}{\vec{r}_{0k}^{\,\,2}}\right)
\ln\left(  \frac{\vec{r}_{jk}^{\,\,2}}{\vec{r}_{0j}^{\,\,2}}\right)
,\quad\int d\vec{r}_{5}J_{ijj05}=0, \label{ints1}%
\end{equation}
one could get the other half of the double logarithm term in the BK kernel
\begin{equation}
\Delta K\rightarrow{\frac{\alpha_{s}^{2}N_{c}}{8\pi^{3}}}\!\int\!d\vec{r}%
_{0}\left(  N_{c}\mathbf{U}_{2^{\dag}1}-\mathbf{U}_{0^{\dag}1}\mathbf{U}%
_{2^{\dag}0}\right)  \frac{\vec{r}_{12}{}^{2}}{\vec{r}_{10}{}^{2}\vec{r}%
_{20}{}^{2}}\ln\left(  \frac{\vec{r}_{12}{}^{2}}{\vec{r}_{10}{}^{2}}\right)
\ln\left(  \frac{\vec{r}_{12}{}^{2}}{\vec{r}_{20}{}^{2}}\right)  .
\label{step-1}%
\end{equation}
Although such treatment gives the correct result, it does not take into
account the IR singularity of $J.$ Indeed if one introduces the dimensional
regularization into (\ref{deltaK}) then one gets%
\begin{align}
\Delta K  &  \rightarrow{\frac{\alpha_{s}^{2}N_{c}}{8\pi^{4}}}\!\int
\!d^{d}r_{0}d^{d}r_{5}\left(  N_{c}\mathbf{U}_{2^{\dag}1}-\mathbf{U}_{0^{\dag
}1}\mathbf{U}_{2^{\dag}0}\right) \nonumber\\
&  \times(J_{22105}-J_{21205}+J_{11205}-J_{12105}+J_{21105}+J_{12205}).
\end{align}
However, in the dimensional regularization the integral $\int d^{d}%
r_{5}J_{ijj05}$ would be $\sim\epsilon$ rather than $0$ and the double
integral
\begin{equation}
\int\!d^{d}r_{0}d^{d}r_{5}J_{ijj05}=2\pi^{2}\zeta(3)
\end{equation}
because the second integral w.r.t. $r_{0}$ has\ an IR divergence as
$r_{0}\rightarrow\infty$ and starts from $\frac{1}{\epsilon}.$ Therefore if
one wants to integrate the coefficient of $\mathbf{U}_{0^{\dag}1}%
\mathbf{U}_{2^{\dag}0}$ w.r.t. $d^{d}r_{5},$ one has to keep the result in the
dimension $d$ without expanding the series. At the same time the coefficient
of $N_{c}\mathbf{U}_{2^{\dag}1}$ gets the doubled contribution since
\begin{equation}
\int\!d\vec{r}_{0}\frac{\vec{r}_{12}{}^{2}}{\vec{r}_{10}{}^{2}\vec{r}_{20}%
{}^{2}}\ln\left(  \frac{\vec{r}_{12}{}^{2}}{\vec{r}_{10}{}^{2}}\right)
\ln\left(  \frac{\vec{r}_{12}{}^{2}}{\vec{r}_{20}{}^{2}}\right)  =4\pi
^{2}\zeta(3).
\end{equation}
Thus, the result depends on the regularization. Such an ambiguity is the
consequence of the fact that the initial expressions do not have the IR
regularization. To avoid this ambiguity one needs the evolution equations for
Wilson lines (\ref{A2}--\ref{A3}) with the IR regularization. Alternatively,
one can write them in the form where the terms which do not depend on both
$U_{5}^{ab}$ and $U_{0}^{ab}$ are integrated w.r.t. the coordinate of the
other gluon.

In this paper the procedure discussed in (\ref{step1}--\ref{step-1}) is used.
Technically it means that for the terms $\sim$ $U_{5}^{ab}U_{0}^{a^{\prime
}b^{\prime}},$ the gluons are treated equally and the kernel is represented in
the form of symmetrized sum (\ref{step1}). In the terms depending only on
$U_{5}^{ab}$ $\ $or only on $U_{0}^{ab},$ the integration order is fixed as
$\int d\vec{r}_{0}\int d\vec{r}_{5}$ or $\int d\vec{r}_{5}\int d\vec{r}_{0}$
correspondingly and the integrals are understood as iterated. As a result, one
can take the inner integral via (\ref{ints1}). The terms independent of
$U_{5}^{ab}$ and $U_{0}^{ab}$ are also symmetrized according to (\ref{step1})
and in them the substitution $J_{ijj05}\rightarrow J_{jji05}$ is made. This
substitution can be understood as follows. First one drops the terms with
$\int d\vec{r}_{5}J_{ijj05}.$ They vanish (\ref{ints1}) if one treats the
integrals as iterated. Next, one adds the totally antisymmetric w.r.t.
$(5\leftrightarrow0)$\ terms $J_{jji05}.$ These terms vanish if they are
integrated w.r.t. $\vec{r}_{0}$ and $\vec{r}_{5}$ in the double integral.
After that the first integral in (\ref{ints1}) is enough to calculate all the
integrals. Again, I stress that although such treatment gives the correct
dipole result (as well as the evolution equation for the baryon operator
coinciding with \cite{bg}) it involves the cavalier treatment of the IR singularities.

Taking the contributions of the self-interaction of one Wilson line
(\ref{Uinitial}), the connected contributions of 2 Wilson lines
(\ref{UUinitial}) and the connected contributions of 3 Wilson lines
(\ref{UUUinitial}) with the appropriate charge conjugation, and using the
integration procedure described above, one can write the full NLO evolution of
the quadrupole operator $tr\left(  U_{1}U_{2}{}^{\dag}U_{3}U_{4}{}^{\dag
}\right)  \equiv\mathbf{U}_{12^{\dag}34^{\dag}}$\ as%
\begin{equation}
\langle K_{NLO}\otimes\mathbf{U}_{12^{\dag}34^{\dag}}\rangle={\frac{\alpha
_{s}^{2}}{8\pi^{4}}}\!\int\!d\vec{r}_{0}d\vec{r}_{5}~\mathbf{(\mathbf{G}}%
_{s}\mathbf{\mathbf{+G}}_{a}\mathbf{)+}{\frac{\alpha_{s}^{2}}{8\pi^{3}}}%
\!\int\!d\vec{r}_{0}~(\mathbf{G}_{\beta}+\mathbf{G)}, \label{Gdefinition}%
\end{equation}
Following (\ref{step1}) the 2-gluon contribution can be decomposed into the
product $\mathbf{G}_{s}$ of the symmetric coordinate and color structures and
the product $\mathbf{G}_{a}$ of the antisymmetric ones w.r.t.
$0\leftrightarrow5$ transposition, i.e. the substitution $\vec{r}%
_{0}\leftrightarrow\vec{r}_{5}$ and $U_{0}\leftrightarrow U_{5}.$ After color
convolution and integration w.r.t. $\vec{r}_{0}$ or $\vec{r}_{5}$ of the
contributions which do not depend on the other variable one comes to the
1-gluon part. It contains the contribution proportional to $\beta$-function
$\mathbf{G}_{\beta}$ ($\beta=\frac{11}{3}-\frac{2}{3}\frac{n_{f}}{N_{c}}$) and
the rest $\mathbf{G}$. One can see that all the $\mathbf{G}$'s separately
vanish without the shockwave, i.e. if all the $U\rightarrow0.$

Doing the same for the double dipole operator $tr\left(  U_{1}U_{2}{}^{\dag
})tr(U_{3}U_{4}{}^{\dag}\right)  \equiv\mathbf{U}_{12^{\dag}}\mathbf{U}%
_{34^{\dag}}$, one can write its full NLO evolution equation\ as%
\begin{align}
\langle K_{NLO}\otimes\mathbf{U}_{12^{\dag}}\mathbf{U}_{34^{\dag}}\rangle &
=\mathbf{U}_{12^{\dag}}\langle K_{NLO}\otimes\mathbf{U}_{34^{\dag}}%
\rangle+\mathbf{U}_{34^{\dag}}\langle K_{NLO}\otimes\mathbf{U}_{12^{\dag}%
}\rangle\nonumber\\
&  +{\frac{\alpha_{s}^{2}}{8\pi^{4}}}\!\int\!d\vec{r}_{0}d\vec{r}%
_{5}~\mathbf{(\mathbf{\tilde{G}}}_{s}\mathbf{\mathbf{+\tilde{G}}}%
_{a}\mathbf{)+}{\frac{\alpha_{s}^{2}}{8\pi^{3}}}\!\int\!d\vec{r}%
_{0}~(\mathbf{\tilde{G}}_{\beta}+\mathbf{\tilde{G})}. \label{Gtildedefinition}%
\end{align}
Here the NLO dipole kernel is written in our notation in (\ref{NLO_BK}),
$\mathbf{\tilde{G}}_{s}(\mathbf{\tilde{G}}_{a})$ is the product of the
coordinate and color structures (anti)symmetric w.r.t. $0\leftrightarrow5$
transposition, $\mathbf{\tilde{G}}_{\beta}$ is proportional to $\beta
$-function and $\mathbf{\tilde{G}}$ is the remaining contribution with 1 gluon
crossing the shockwave.

\subsection{Quadrupole}

We start from the product of the symmetric structures%
\begin{equation}
\mathbf{G}_{s}=\mathbf{G}_{s1}+n_{f}\mathbf{G}_{q}+\mathbf{G}_{s2}.
\end{equation}%
\begin{align}
\mathbf{G}_{s1}=  &  \left(  \left\{  \mathbf{U}_{0^{\dag}34^{\dag}15^{\dag
}02^{\dag}5}-\mathbf{U}_{5^{\dag}0}\mathbf{U}_{2^{\dag}5}\mathbf{U}_{0^{\dag
}34^{\dag}1}-(5\rightarrow0)\right\}  +(5\leftrightarrow0)\right)  \left(
L_{12}+L_{32}-L_{13}\right) \nonumber\\
+  &  \left(  \left\{  \mathbf{U}_{0^{\dag}15^{\dag}02^{\dag}34^{\dag}%
5}-\mathbf{U}_{0^{\dag}5}\mathbf{U}_{5^{\dag}1}\mathbf{U}_{2^{\dag}34^{\dag}%
0}-(5\rightarrow0)\right\}  +(5\leftrightarrow0)\right)  (L_{12}+L_{14}%
-L_{42})\nonumber\\
+  &  (1\leftrightarrow3,2\leftrightarrow4),
\end{align}
where $L$ was introduced in (\ref{L12}). It is a conformally invariant
contribution.%
\begin{align}
\mathbf{G}_{q}=  &  (\{\frac{\mathbf{U}_{0^{\dag}34^{\dag}12^{\dag}%
5}+\mathbf{U}_{2^{\dag}34^{\dag}15^{\dag}0}}{N_{c}}-\frac{\mathbf{U}_{0^{\dag
}5}\mathbf{U}_{2^{\dag}34^{\dag}1}}{N_{c}^{2}}-\mathbf{U}_{2^{\dag}%
5}\mathbf{U}_{0^{\dag}34^{\dag}1}-(5\rightarrow0)\}+(5\leftrightarrow
0))\nonumber\\
\times &  \frac{1}{2}\left(  L_{12}^{q}+L_{32}^{q}-L_{13}^{q}\right)
+\frac{1}{2}\left(  L_{12}^{q}+L_{14}^{q}-L_{42}^{q}\right) \nonumber\\
\times &  (\{\frac{\mathbf{U}_{0^{\dag}12^{\dag}34^{\dag}5}+\mathbf{U}%
_{2^{\dag}34^{\dag}15^{\dag}0}}{N_{c}}-\frac{\mathbf{U}_{0^{\dag}5}%
\mathbf{U}_{2^{\dag}34^{\dag}1}}{N_{c}^{2}}-\mathbf{U}_{5^{\dag}1}%
\mathbf{U}_{2^{\dag}34^{\dag}0}-(5\rightarrow0)\}+(5\leftrightarrow
0))\nonumber\\
+  &  (1\leftrightarrow3,2\leftrightarrow4), \label{Gq}%
\end{align}
Here $L^{q}$ is defined in (\ref{L12q}).%
\begin{align}
\mathbf{G}_{s2}\mathbf{=}  &  \frac{1}{2}\left(  \mathbf{U}_{0^{\dag}34^{\dag
}52^{\dag}05^{\dag}1}-\mathbf{U}_{0^{\dag}1}\mathbf{U}_{2^{\dag}5}%
\mathbf{U}_{4^{\dag}05^{\dag}3}+(5\leftrightarrow0)\right)  (M_{1}^{34}%
-M_{1}^{24}+M_{2}^{43}-M_{2}^{13}+(5\leftrightarrow0))\nonumber\\
+  &  \frac{1}{2}\left(  \mathbf{U}_{0^{\dag}35^{\dag}02^{\dag}54^{\dag}%
1}-\mathbf{U}_{0^{\dag}3}\mathbf{U}_{2^{\dag}5}\mathbf{U}_{4^{\dag}15^{\dag}%
0}+(5\leftrightarrow0)\right)  (M_{3}^{14}-M_{3}^{24}+M_{2}^{41}-M_{2}%
^{31}+(5\leftrightarrow0))\nonumber\\
+  &  \frac{1}{2}\left(  \mathbf{U}_{0^{\dag}15^{\dag}02^{\dag}34^{\dag}%
5}-\mathbf{U}_{0^{\dag}5}\mathbf{U}_{5^{\dag}1}\mathbf{U}_{2^{\dag}34^{\dag}%
0}+(5\leftrightarrow0)\right)  (M_{2}^{14}+M_{4}^{12}+(5\leftrightarrow
0))\nonumber\\
+  &  \frac{1}{2}\left(  \mathbf{U}_{0^{\dag}34^{\dag}15^{\dag}02^{\dag}%
5}-\mathbf{U}_{5^{\dag}0}\mathbf{U}_{2^{\dag}5}\mathbf{U}_{0^{\dag}34^{\dag}%
1}+(5\leftrightarrow0)\right)  (M_{1}^{23}+M_{3}^{21}+(5\leftrightarrow
0))\nonumber\\
+  &  (1\leftrightarrow3,2\leftrightarrow4),
\end{align}
where $M_{i}^{jk}$ is defined in (\ref{M}). Using property
(\ref{Masimproperty}) one can show that $\mathbf{G}_{s2}$\ vanishes without
the shockwave, i.e. when all the $U\rightarrow1.$ Indeed, it is clear from the
representation%
\[
2\mathbf{G}_{s2}=(\mathbf{U}_{2^{\dagger}5}\mathbf{U}_{5^{\dagger}0}%
\mathbf{U}_{0^{\dagger}34^{\dagger}1}-\mathbf{U}_{4^{\dagger}0}\mathbf{U}%
_{5^{\dagger}1}\mathbf{U}_{0^{\dagger}52^{\dagger}3}+\mathbf{U}_{0^{\dagger
}15^{\dagger}04^{\dagger}52^{\dagger}3}-\mathbf{U}_{0^{\dagger}34^{\dagger
}15^{\dagger}02^{\dagger}5}+(5\leftrightarrow0))
\]%
\[
\times(M_{1}^{32}-M_{1}^{23})
\]%
\[
+(\mathbf{U}_{4^{\dagger}0}\mathbf{U}_{5^{\dagger}1}\mathbf{U}_{0^{\dagger
}52^{\dagger}3}-\mathbf{U}_{2^{\dagger}0}\mathbf{U}_{5^{\dagger}1}%
\mathbf{U}_{0^{\dagger}34^{\dagger}5}-\mathbf{U}_{0^{\dagger}15^{\dagger
}04^{\dagger}52^{\dagger}3}+\mathbf{U}_{0^{\dagger}15^{\dagger}34^{\dagger
}02^{\dagger}5}+(5\leftrightarrow0))(M_{1}^{42}-M_{1}^{24})
\]%
\[
+(\mathbf{U}_{0^{\dagger}1}\mathbf{U}_{2^{\dagger}5}\mathbf{U}_{4^{\dagger
}05^{\dagger}3}-\mathbf{U}_{4^{\dagger}5}\mathbf{U}_{5^{\dagger}0}%
\mathbf{U}_{0^{\dagger}12^{\dagger}3}+\mathbf{U}_{0^{\dagger}12^{\dagger
}35^{\dagger}04^{\dagger}5}-\mathbf{U}_{0^{\dagger}15^{\dagger}34^{\dagger
}02^{\dagger}5}+(5\leftrightarrow0))(M_{1}^{43}-M_{1}^{34})
\]%
\[
+(\mathbf{U}_{0^{\dagger}3}\mathbf{U}_{2^{\dagger}5}\mathbf{U}_{4^{\dagger
}15^{\dagger}0}-\mathbf{U}_{2^{\dagger}0}\mathbf{U}_{5^{\dagger}1}%
\mathbf{U}_{0^{\dagger}34^{\dagger}5}+\mathbf{U}_{0^{\dagger}15^{\dagger
}34^{\dagger}02^{\dagger}5}-\mathbf{U}_{0^{\dagger}52^{\dagger}04^{\dagger
}15^{\dagger}3}+(5\leftrightarrow0))(M_{2}^{31}-M_{2}^{13})
\]%
\[
+(\mathbf{U}_{0^{\dagger}5}\mathbf{U}_{5^{\dagger}1}\mathbf{U}_{2^{\dagger
}34^{\dagger}0}-\mathbf{U}_{0^{\dagger}3}\mathbf{U}_{2^{\dagger}5}%
\mathbf{U}_{4^{\dagger}15^{\dagger}0}-\mathbf{U}_{0^{\dagger}15^{\dagger
}02^{\dagger}34^{\dagger}5}+\mathbf{U}_{0^{\dagger}35^{\dagger}02^{\dagger
}54^{\dagger}1}+(5\leftrightarrow0))({}M_{2}^{41}-M_{2}^{14})
\]%
\[
+(\mathbf{U}_{0^{\dagger}5}\mathbf{U}_{5^{\dagger}3}\mathbf{U}_{2^{\dagger
}04^{\dagger}1}-\mathbf{U}_{2^{\dagger}0}\mathbf{U}_{5^{\dagger}1}%
\mathbf{U}_{0^{\dagger}34^{\dagger}5}+\mathbf{U}_{0^{\dagger}15^{\dagger
}34^{\dagger}02^{\dagger}5}-\mathbf{U}_{0^{\dagger}54^{\dagger}12^{\dagger
}05^{\dagger}3}+(5\leftrightarrow0))(M_{2}^{43}-M_{2}^{34})
\]%
\begin{equation}
+(1\leftrightarrow3,2\leftrightarrow4). \label{Gs2vanishproof}%
\end{equation}
The contribution which is the product of the antisymmetric w.r.t.
$5\leftrightarrow0$ parts reads
\begin{equation}
\mathbf{G}_{a}=\mathbf{G}_{a1}+\mathbf{G}_{a2}+\mathbf{G}_{a3}.
\end{equation}%
\begin{align}
\mathbf{G}_{a1}\mathbf{=}  &  \frac{1}{2}\left(  \mathbf{U}_{0^{\dag}%
1}\mathbf{U}_{2^{\dag}5}\mathbf{U}_{4^{\dag}05^{\dag}3}+\mathbf{U}_{0^{\dag
}34^{\dag}52^{\dag}05^{\dag}1}-(5\leftrightarrow0)\right)  (M_{2}^{31}%
-M_{2}^{34}-M_{1}^{42}+M_{1}^{43}-(5\leftrightarrow0))\nonumber\\
+  &  \frac{1}{2}\left(  \mathbf{U}_{0^{\dag}3}\mathbf{U}_{2^{\dag}%
5}\mathbf{U}_{4^{\dag}15^{\dag}0}+\mathbf{U}_{0^{\dag}35^{\dag}02^{\dag
}54^{\dag}1}-(5\leftrightarrow0)\right)  (M_{2}^{13}-M_{2}^{14}-M_{3}%
^{42}+M_{3}^{41}-(5\leftrightarrow0))\nonumber\\
+  &  (1\leftrightarrow3,2\leftrightarrow4).
\end{align}%
\begin{align}
\mathbf{G}_{a2}\mathbf{=}  &  \frac{1}{2}\left(  \mathbf{U}_{0^{\dag}34^{\dag
}15^{\dag}02^{\dag}5}-(5\leftrightarrow0)\right)  (\tilde{L}_{13}%
+2M_{21}-2M_{23}-M_{1}^{23}+M_{3}^{21}-(5\leftrightarrow0))\nonumber\\
+  &  \frac{1}{2}\left(  \mathbf{U}_{0^{\dag}15^{\dag}02^{\dag}34^{\dag}%
5}-(5\leftrightarrow0)\right)  (\tilde{L}_{42}-2M_{12}+2M_{14}+M_{2}%
^{14}-M_{4}^{12}-(5\leftrightarrow0))\nonumber\\
+  &  (1\leftrightarrow3,2\leftrightarrow4).
\end{align}
Here the functions $\tilde{L}$ and $M_{ij}$ are defined in (\ref{Ltilde}) and
(\ref{Mij}).%
\begin{align}
\mathbf{G}_{a3}\mathbf{=}  &  \frac{1}{2}\left(  \mathbf{U}_{0^{\dag}%
5}\mathbf{U}_{5^{\dag}1}\mathbf{U}_{2^{\dag}34^{\dag}0}-(5\leftrightarrow
0)\right)  (\tilde{L}_{12}+\tilde{L}_{14}-2M_{24}+M_{2}^{14}+M_{4}%
^{12}-(5\leftrightarrow0))\nonumber\\
+  &  \frac{1}{2}\left(  \mathbf{U}_{5^{\dag}0}\mathbf{U}_{2^{\dag}%
5}\mathbf{U}_{0^{\dag}34^{\dag}1}-(5\leftrightarrow0)\right)  (\tilde{L}%
_{21}+\tilde{L}_{23}-2M_{13}+M_{1}^{23}+M_{3}^{21}-(5\leftrightarrow
0))\nonumber\\
+  &  (1\leftrightarrow3,2\leftrightarrow4).
\end{align}
From (\ref{Ltilde}) and (\ref{Mij}) one can see that it is possible to express
$\mathbf{G}_{a}$ in terms of only one function $M_{i}^{jk}$ (\ref{M}).

The $\beta$-functional part of 1-gluon contribution $\mathbf{G}_{\beta}$
(\ref{Gdefinition}) has the same structure as LO kernel (\ref{LOkernel})%
\begin{align}
\mathbf{G}_{\beta}  &  =\frac{\vec{r}_{14}{}^{2}}{\vec{r}_{10}{}^{2}\vec
{r}_{40}{}^{2}}M_{14}^{\beta}\left(  \mathbf{U}_{10^{\dag}}\mathbf{U}%
_{02^{\dag}34^{\dag}}+\mathbf{U}_{4^{\dag}0}\mathbf{U}_{12^{\dag}30^{\dag}%
}-\left(  0\rightarrow1\equiv0\rightarrow4\right)  \right) \nonumber\\
+  &  \frac{\vec{r}_{12}{}^{2}}{\vec{r}_{10}{}^{2}\vec{r}_{20}{}^{2}}%
M_{12}^{\beta}\left(  \mathbf{U}_{10^{\dag}}\mathbf{U}_{02^{\dag}34^{\dag}%
}+\mathbf{U}_{2^{\dag}0}\mathbf{U}_{10^{\dag}34^{\dag}}-\left(  0\rightarrow
1\equiv0\rightarrow2\right)  \right) \nonumber\\
-  &  \frac{\vec{r}_{24}{}^{2}}{2\vec{r}_{20}{}^{2}\vec{r}_{40}{}^{2}}%
M_{24}^{\beta}\left(  \mathbf{U}_{10^{\dag}}\mathbf{U}_{02^{\dag}34^{\dag}%
}+\mathbf{U}_{30^{\dag}}\mathbf{U}_{04^{\dag}12^{\dag}}-(0\rightarrow
4\equiv0\rightarrow2)\right) \nonumber\\
-  &  \frac{\vec{r}_{13}{}^{2}}{2\vec{r}_{10}{}^{2}\vec{r}_{30}{}^{2}}%
M_{13}^{\beta}\left(  \mathbf{U}_{4^{\dag}0}\mathbf{U}_{12^{\dag}30^{\dag}%
}+\mathbf{U}_{2^{\dag}0}\mathbf{U}_{34^{\dag}10^{\dag}}-(0\rightarrow
1\equiv0\rightarrow3)\right)  +(1\leftrightarrow3,2\leftrightarrow4).
\label{beta-term}%
\end{align}
Here $M^{\beta}$ is defined in (\ref{Mbeta}). The 1-gluon term without beta
function reads%
\begin{equation}
\mathbf{G=G}_{1}+\mathbf{G}_{0}.
\end{equation}
In $\mathbf{G}$ one can pick the terms independent of $\vec{r}_{0}$ and
integrate them out if they are convergent. We call these terms $\mathbf{G}%
_{0}.$ In fact the choice of $\mathbf{G}_{0}$ is not unique. We have%
\begin{align}
\mathbf{G}_{0}=  &  \frac{N_{c}}{4}(\mathbf{U}_{4^{\dag}1}\mathbf{U}_{2^{\dag
}3}-\mathbf{U}_{4^{\dag}3}\mathbf{U}_{2^{\dag}1})\left\{  \left(  \frac
{\vec{r}_{14}{}^{2}}{\vec{r}_{10}{}^{2}\vec{r}_{40}{}^{2}}+\frac{\vec{r}%
_{23}{}^{2}}{\vec{r}_{20}{}^{2}\vec{r}_{30}{}^{2}}-\frac{\vec{r}_{13}{}^{2}%
}{\vec{r}_{10}{}^{2}\vec{r}_{30}{}^{2}}-\frac{\vec{r}_{24}{}^{2}}{\vec{r}%
_{20}{}^{2}\vec{r}_{40}{}^{2}}\right)  \right. \nonumber\\
\times &  \ln\left(  \frac{\vec{r}_{10}{}^{2}}{\vec{r}_{12}{}^{2}}\right)
\ln\left(  \frac{\vec{r}_{20}{}^{2}}{\vec{r}_{12}{}^{2}}\right)  +\left(
\frac{\vec{r}_{13}{}^{2}}{\vec{r}_{10}{}^{2}\vec{r}_{30}{}^{2}}+\frac{\vec
{r}_{24}{}^{2}}{\vec{r}_{20}{}^{2}\vec{r}_{40}{}^{2}}-\frac{\vec{r}_{34}{}%
^{2}}{\vec{r}_{30}{}^{2}\vec{r}_{40}{}^{2}}-\frac{\vec{r}_{12}{}^{2}}{\vec
{r}_{20}{}^{2}\vec{r}_{10}{}^{2}}\right) \nonumber\\
\times &  \ln\left(  \frac{\vec{r}_{10}{}^{2}}{\vec{r}_{14}{}^{2}}\right)
\ln\left(  \frac{\vec{r}_{40}{}^{2}}{\vec{r}_{14}{}^{2}}\right)  +\left(
\ln\left(  \frac{\vec{r}_{20}{}^{2}}{\vec{r}_{24}{}^{2}}\right)  \ln\left(
\frac{\vec{r}_{40}{}^{2}}{\vec{r}_{24}{}^{2}}\right)  +\ln\left(  \frac
{\vec{r}_{10}{}^{2}}{\vec{r}_{13}{}^{2}}\right)  \ln\left(  \frac{\vec{r}%
_{30}{}^{2}}{\vec{r}_{13}{}^{2}}\right)  \right) \nonumber\\
\times &  \left.  \left(  \frac{\vec{r}_{12}{}^{2}}{\vec{r}_{10}{}^{2}\vec
{r}_{20}{}^{2}}-\frac{\vec{r}_{14}{}^{2}}{\vec{r}_{10}{}^{2}\vec{r}_{40}{}%
^{2}}\right)  \right\}  +(1\leftrightarrow3,2\leftrightarrow4).
\end{align}
It has zero dipole limit.%
\begin{align*}
\mathbf{G=}  &  \frac{\vec{r}_{12}{}^{2}}{\vec{r}_{10}{}^{2}\vec{r}_{20}{}%
^{2}}\ln\left(  \frac{\vec{r}_{10}{}^{2}}{\vec{r}_{12}{}^{2}}\right)
\ln\left(  \frac{\vec{r}_{20}{}^{2}}{\vec{r}_{12}{}^{2}}\right)  \{\frac
{N_{c}}{2}\left(  2N_{c}\mathbf{U}_{2^{\dag}34^{\dag}1}-\mathbf{U}_{0^{\dag}%
1}\mathbf{U}_{2^{\dag}34^{\dag}0}-\mathbf{U}_{2^{\dag}0}\mathbf{U}_{4^{\dag
}10^{\dag}3}\right) \\
+  &  \left(  \mathbf{U}_{2^{\dag}10^{\dag}34^{\dag}0}-\mathbf{U}_{2^{\dag}%
0}\mathbf{U}_{4^{\dag}3}\mathbf{U}_{0^{\dag}1}-(0\rightarrow1)\right)  \}\\
+  &  \frac{\vec{r}_{14}{}^{2}}{\vec{r}_{10}{}^{2}\vec{r}_{40}{}^{2}}%
\ln\left(  \frac{\vec{r}_{10}{}^{2}}{\vec{r}_{14}{}^{2}}\right)  \ln\left(
\frac{\vec{r}_{40}{}^{2}}{\vec{r}_{14}{}^{2}}\right)  \{\frac{N_{c}}{2}\left(
2N_{c}\mathbf{U}_{2^{\dag}34^{\dag}1}-\mathbf{U}_{0^{\dag}1}\mathbf{U}%
_{2^{\dag}34^{\dag}0}-\mathbf{U}_{4^{\dag}0}\mathbf{U}_{2^{\dag}30^{\dag}%
1}\right) \\
+  &  \left(  \mathbf{U}_{2^{\dag}30^{\dag}14^{\dag}0}-\mathbf{U}_{4^{\dag}%
0}\mathbf{U}_{2^{\dag}3}\mathbf{U}_{0^{\dag}1}-(0\rightarrow1)\right)  \}\\
+  &  \frac{1}{2}\left(  \frac{\vec{r}_{13}{}^{2}}{\vec{r}_{10}{}^{2}\vec
{r}_{30}{}^{2}}\ln\left(  \frac{\vec{r}_{10}{}^{2}}{\vec{r}_{13}{}^{2}%
}\right)  \ln\left(  \frac{\vec{r}_{30}{}^{2}}{\vec{r}_{13}{}^{2}}\right)
+\frac{\vec{r}_{24}{}^{2}}{\vec{r}_{20}{}^{2}\vec{r}_{40}{}^{2}}\ln\left(
\frac{\vec{r}_{20}{}^{2}}{\vec{r}_{24}{}^{2}}\right)  \ln\left(  \frac{\vec
{r}_{40}{}^{2}}{\vec{r}_{24}{}^{2}}\right)  \right) \\
\times &  \left\{  (\mathbf{U}_{4^{\dag}0}\mathbf{U}_{2^{\dag}1}%
+\mathbf{U}_{4^{\dag}1}\mathbf{U}_{2^{\dag}0})\mathbf{U}_{0^{\dag}%
3}-\mathbf{U}_{2^{\dag}04^{\dag}10^{\dag}3}-\mathbf{U}_{2^{\dag}04^{\dag
}30^{\dag}1}-(0\rightarrow3)\right\}
\end{align*}%
\begin{align*}
&  +\left\{  \mathbf{U}_{2^{\dag}0}\mathbf{U}_{4^{\dag}1}\mathbf{U}_{0^{\dag
}3}-\mathbf{U}_{2^{\dag}0}\mathbf{U}_{0^{\dag}1}\mathbf{U}_{34^{\dag}%
}+\mathbf{U}_{2^{\dag}10^{\dag}34^{\dag}0}-\mathbf{U}_{0^{\dag}32^{\dag
}04^{\dag}1}\right\} \\
&  \times\frac{1}{2\vec{r}_{20}{}^{2}}\left(  \frac{\vec{r}_{23}{}^{2}}%
{\vec{r}_{30}{}^{2}}-\frac{\vec{r}_{12}{}^{2}}{\vec{r}_{10}{}^{2}}\right)
\ln\left(  \frac{\vec{r}_{10}{}^{2}}{\vec{r}_{13}{}^{2}}\right)  \ln\left(
\frac{\vec{r}_{30}{}^{2}}{\vec{r}_{13}{}^{2}}\right)  +\frac{1}{2\vec{r}%
_{10}{}^{2}}\left(  \frac{\vec{r}_{14}{}^{2}}{\vec{r}_{40}{}^{2}}-\frac
{\vec{r}_{12}{}^{2}}{\vec{r}_{20}{}^{2}}\right)  \ln\left(  \frac{\vec{r}%
_{20}{}^{2}}{\vec{r}_{24}{}^{2}}\right)  \ln\left(  \frac{\vec{r}_{40}{}^{2}%
}{\vec{r}_{24}{}^{2}}\right) \\
&  \times\left\{  \mathbf{U}_{2^{\dag}3}\mathbf{U}_{4^{\dag}0}\mathbf{U}%
_{0^{\dag}1}-\mathbf{U}_{2^{\dag}0}\mathbf{U}_{0^{\dag}1}\mathbf{U}_{34^{\dag
}}+\mathbf{U}_{2^{\dag}10^{\dag}34^{\dag}0}-\mathbf{U}_{0^{\dag}14^{\dag
}02^{\dag}3}\right\}
\end{align*}%
\begin{align}
&  +\left\{  \mathbf{U}_{4^{\dag}0}\mathbf{U}_{2^{\dag}1}\mathbf{U}_{0^{\dag
}3}-N_{c}\mathbf{U}_{2^{\dag}0}\mathbf{U}_{4^{\dag}10^{\dag}3}+\mathbf{U}%
_{2^{\dag}34^{\dag}1}-\mathbf{U}_{2^{\dag}04^{\dag}30^{\dag}1}\right\}
\nonumber\\
&  \times\frac{1}{2\vec{r}_{30}{}^{2}}\left(  \frac{\vec{r}_{23}{}^{2}}%
{\vec{r}_{20}{}^{2}}-\frac{\vec{r}_{13}{}^{2}}{\vec{r}_{10}{}^{2}}\right)
\ln\left(  \frac{\vec{r}_{10}{}^{2}}{\vec{r}_{12}{}^{2}}\right)  \ln\left(
\frac{\vec{r}_{20}{}^{2}}{\vec{r}_{12}{}^{2}}\right)  +\frac{1}{2\vec{r}%
_{40}{}^{2}}\left(  \frac{\vec{r}_{14}{}^{2}}{\vec{r}_{10}{}^{2}}-\frac
{\vec{r}_{24}{}^{2}}{\vec{r}_{20}{}^{2}}\right)  \ln\left(  \frac{\vec{r}%
_{10}{}^{2}}{\vec{r}_{12}{}^{2}}\right)  \ln\left(  \frac{\vec{r}_{20}{}^{2}%
}{\vec{r}_{12}{}^{2}}\right) \nonumber\\
&  \times\left\{  \mathbf{U}_{4^{\dag}0}\mathbf{U}_{2^{\dag}1}\mathbf{U}%
_{0^{\dag}3}-N_{c}\mathbf{U}_{0^{\dag}1}\mathbf{U}_{2^{\dag}34^{\dag}%
0}+\mathbf{U}_{2^{\dag}34^{\dag}1}-\mathbf{U}_{2^{\dag}04^{\dag}30^{\dag}%
1}\right\} \nonumber
\end{align}%
\begin{align}
&  +\left\{  \mathbf{U}_{2^{\dag}0}\mathbf{U}_{4^{\dag}1}\mathbf{U}_{0^{\dag
}3}-N_{c}\mathbf{U}_{4^{\dag}0}\mathbf{U}_{12^{\dag}30^{\dag}}+\mathbf{U}%
_{2^{\dag}34^{\dag}1}-\mathbf{U}_{2^{\dag}04^{\dag}10^{\dag}3}\right\}
\nonumber\\
&  \times\frac{1}{2\vec{r}_{30}{}^{2}}\left(  \frac{\vec{r}_{34}{}^{2}}%
{\vec{r}_{40}{}^{2}}-\frac{\vec{r}_{13}{}^{2}}{\vec{r}_{10}{}^{2}}\right)
\ln\left(  \frac{\vec{r}_{10}{}^{2}}{\vec{r}_{14}{}^{2}}\right)  \ln\left(
\frac{\vec{r}_{40}{}^{2}}{\vec{r}_{14}{}^{2}}\right)  +\frac{1}{2\vec{r}%
_{20}{}^{2}}\left(  \frac{\vec{r}_{12}{}^{2}}{\vec{r}_{10}{}^{2}}-\frac
{\vec{r}_{24}{}^{2}}{\vec{r}_{40}{}^{2}}\right)  \ln\left(  \frac{\vec{r}%
_{10}{}^{2}}{\vec{r}_{14}{}^{2}}\right)  \ln\left(  \frac{\vec{r}_{40}{}^{2}%
}{\vec{r}_{14}{}^{2}}\right) \nonumber\\
&  \times\left\{  \mathbf{U}_{2^{\dag}0}\mathbf{U}_{4^{\dag}1}\mathbf{U}%
_{0^{\dag}3}-N_{c}\mathbf{U}_{0^{\dag}1}\mathbf{U}_{02^{\dag}34^{\dag}%
}+\mathbf{U}_{2^{\dag}34^{\dag}1}-\mathbf{U}_{2^{\dag}04^{\dag}10^{\dag}%
3}\right\}  +(1\leftrightarrow3,2\leftrightarrow4).
\end{align}
All the integrals with the functions $\mathbf{G}_{s},$ $\mathbf{G}_{a},$
$\mathbf{G}_{\beta}$ and $\mathbf{G}$ are convergent. It is clear from the
explicit expressions for $\mathbf{G}_{\beta}$ and $\mathbf{G}.$ For
$\mathbf{G}_{s}$ and $\mathbf{G}_{a}$ one can see it recalling that
$L_{ij}^{(q)}$ has unintegrable singularity at $\vec{r}_{0}=\vec{r}_{5}$ and
$M_{k}^{ij}$ has unintegrable singularity at $\vec{r}_{0}=\vec{r}_{5}=\vec
{r}_{k}.$ In all expressions in this section these singularities cancel.

\subsection{Double dipole}

The symmetric contribution reads%
\begin{equation}
\mathbf{\tilde{G}}_{s}=\mathbf{\tilde{G}}_{s1}+n_{f}\mathbf{\tilde{G}}%
_{q}+\mathbf{\tilde{G}}_{s2},
\end{equation}%
\begin{align}
\mathbf{\tilde{G}}_{s1}=  &  \left(  \left\{  \mathbf{U}_{0^{\dag}12^{\dag}%
5}\mathbf{U}_{4^{\dag}35^{\dag}0}-\mathbf{U}_{0^{\dag}5}\mathbf{U}_{2^{\dag
}15^{\dag}34^{\dag}0}-(5\rightarrow0)\right\}  +(5\leftrightarrow0)\right)
\left(  L_{14}-L_{13}+L_{23}-L_{24}\right) \nonumber\\
+  &  (1\leftrightarrow3,2\leftrightarrow4),\\
\mathbf{\tilde{G}}_{q}=  &  \frac{1}{2}(\{\frac{\mathbf{U}_{4^{\dag}3}}{N_{c}%
}(\mathbf{U}_{0^{\dag}12^{\dag}5}+\mathbf{U}_{0^{\dag}52^{\dag}1}%
-\frac{\mathbf{U}_{2^{\dag}1}\mathbf{U}_{0^{\dag}5}}{N_{c}})-\mathbf{U}%
_{0^{\dag}12^{\dag}54^{\dag}3}-(5\rightarrow0)\}+(5\leftrightarrow
0))\nonumber\\
\times &  \left(  L_{14}^{q}-L_{13}^{q}+L_{23}^{q}-L_{24}^{q}\right)
+(1\leftrightarrow3,2\leftrightarrow4),\\
\mathbf{\tilde{G}}_{s2}=  &  \frac{1}{2}(\mathbf{U}_{0^{\dag}54^{\dag}%
3}\mathbf{U}_{2^{\dag}05^{\dag}1}-\mathbf{U}_{0^{\dag}5}\mathbf{U}_{2^{\dag
}15^{\dag}34^{\dag}0}+(5\leftrightarrow0))(M_{4}^{12}+M_{3}^{21}-M_{1}%
^{34}-M_{2}^{43}+(5\leftrightarrow0))\nonumber\\
+  &  \frac{1}{2}(\mathbf{U}_{4^{\dag}0}\mathbf{U}_{0^{\dag}35^{\dag}12^{\dag
}5}+\mathbf{U}_{4^{\dag}0}\mathbf{U}_{0^{\dag}52^{\dag}15^{\dag}3}%
-\mathbf{U}_{0^{\dag}3}\mathbf{U}_{2^{\dag}15^{\dag}04^{\dag}5}-\mathbf{U}%
_{0^{\dag}3}\mathbf{U}_{2^{\dag}54^{\dag}05^{\dag}1}+(5\leftrightarrow
0))\nonumber\\
\times &  \left(  M_{4}^{13}+M_{3}^{14}-M_{4}^{23}-M_{3}^{24}%
+(5\leftrightarrow0)\right)  +(1\leftrightarrow3,2\leftrightarrow4).
\end{align}
Here $L,\,L^{q},$ and $M_{i}^{jk}$ are introduced in (\ref{L12}),
(\ref{L12q}), and (\ref{M}). The antisymmetric contribution reads%
\begin{align}
\mathbf{\mathbf{\tilde{G}}}_{a}=  &  \frac{1}{2}(\mathbf{U}_{0^{\dag}54^{\dag
}3}\mathbf{U}_{2^{\dag}05^{\dag}1}-\mathbf{U}_{0^{\dag}52^{\dag}1}%
\mathbf{U}_{4^{\dag}05^{\dag}3}+\mathbf{U}_{5^{\dag}0}\mathbf{U}_{0^{\dag
}12^{\dag}54^{\dag}3}-\mathbf{U}_{0^{\dag}5}\mathbf{U}_{2^{\dag}15^{\dag
}34^{\dag}0}-(5\leftrightarrow0))\nonumber\\
\times &  (M_{4}^{11}-M_{3}^{11}+M_{3}^{12}-M_{4}^{12}+M_{3}^{21}-M_{4}%
^{21}-M_{3}^{22}+M_{4}^{22})\nonumber\\
+  &  \frac{1}{2}(\mathbf{U}_{5^{\dag}3}\mathbf{U}_{0^{\dag}12^{\dag}04^{\dag
}5}+\mathbf{U}_{5^{\dag}3}\mathbf{U}_{0^{\dag}54^{\dag}02^{\dag}1}%
-\mathbf{U}_{4^{\dag}5}\mathbf{U}_{0^{\dag}12^{\dag}05^{\dag}3}-\mathbf{U}%
_{4^{\dag}5}\mathbf{U}_{0^{\dag}35^{\dag}02^{\dag}1}-(5\leftrightarrow
0))\nonumber\\
\times &  (M_{4}^{23}+M_{3}^{24}-M_{4}^{13}-M_{3}^{14}-M_{4}^{31}+M_{4}%
^{32}-M_{3}^{41}+M_{3}^{42})+(1\leftrightarrow3,2\leftrightarrow4).
\end{align}
The $\beta$-functional contribution has the form%
\begin{align}
\mathbf{\tilde{G}}_{\beta}=  &  \left(  \frac{\vec{r}_{13}{}^{2}}{\vec{r}%
_{10}{}^{2}\vec{r}_{30}{}^{2}}M_{13}^{\beta}-\frac{\vec{r}_{23}{}^{2}}{\vec
{r}_{20}{}^{2}\vec{r}_{30}{}^{2}}M_{23}^{\beta}-\frac{\vec{r}_{14}{}^{2}}%
{\vec{r}_{10}{}^{2}\vec{r}_{40}{}^{2}}M_{14}^{\beta}+\frac{\vec{r}_{24}{}^{2}%
}{\vec{r}_{20}{}^{2}\vec{r}_{40}{}^{2}}M_{24}^{\beta}\right) \nonumber\\
\times &  (\mathbf{U}_{2^{\dag}14^{\dag}3}+\mathbf{U}_{2^{\dag}34^{\dag}%
1}-\mathbf{U}_{2^{\dag}10^{\dag}34^{\dag}0}-\mathbf{U}_{2^{\dag}04^{\dag
}30^{\dag}1}),
\end{align}
where $M^{\beta}$ is introduced in (\ref{Mbeta}). The remaining contribution
reads
\begin{equation}
\mathbf{\tilde{G}=\tilde{G}}_{1}+\mathbf{\tilde{G}}_{0}.
\end{equation}%
\begin{align}
\mathbf{\tilde{G}}_{0}\mathbf{=}  &  \frac{1}{4}(2\mathbf{U}_{2^{\dag}%
1}\mathbf{U}_{4^{\dag}3}-N_{c}\mathbf{U}_{2^{\dag}14^{\dag}3}-N_{c}%
\mathbf{U}_{2^{\dag}34^{\dag}1})\left[  \left(  \frac{2\vec{r}_{13}{}^{2}%
}{\vec{r}_{01}{}^{2}\vec{r}_{03}{}^{2}}-\frac{\vec{r}_{12}{}^{2}}{\vec{r}%
_{01}{}^{2}\vec{r}_{02}{}^{2}}+\frac{\vec{r}_{14}{}^{2}}{\vec{r}_{01}{}%
^{2}\vec{r}_{04}{}^{2}}\right)  \right. \nonumber\\
\times &  \ln\left(  \frac{\vec{r}_{13}{}^{2}}{\vec{r}_{01}{}^{2}}\right)
\ln\left(  \frac{\vec{r}_{13}{}^{2}}{\vec{r}_{03}{}^{2}}\right)  +\left(
\frac{2\vec{r}_{24}{}^{2}}{\vec{r}_{02}{}^{2}\vec{r}_{04}{}^{2}}-\frac{\vec
{r}_{12}{}^{2}}{\vec{r}_{01}{}^{2}\vec{r}_{02}{}^{2}}+\frac{\vec{r}_{14}{}%
^{2}}{\vec{r}_{01}{}^{2}\vec{r}_{04}{}^{2}}\right)  \ln\left(  \frac{\vec
{r}_{24}{}^{2}}{\vec{r}_{02}{}^{2}}\right)  \ln\left(  \frac{\vec{r}_{24}%
{}^{2}}{\vec{r}_{04}{}^{2}}\right) \nonumber\\
+  &  \left(  \frac{\vec{r}_{12}{}^{2}}{\vec{r}_{01}{}^{2}\vec{r}_{02}{}^{2}%
}-\frac{\vec{r}_{13}{}^{2}}{\vec{r}_{01}{}^{2}\vec{r}_{03}{}^{2}}-\frac
{4\vec{r}_{14}{}^{2}}{\vec{r}_{01}{}^{2}\vec{r}_{04}{}^{2}}-\frac{\vec{r}%
_{24}{}^{2}}{\vec{r}_{02}{}^{2}\vec{r}_{04}{}^{2}}+\frac{\vec{r}_{34}{}^{2}%
}{\vec{r}_{03}{}^{2}\vec{r}_{04}{}^{2}}\right)  \ln\left(  \frac{\vec{r}%
_{14}{}^{2}}{\vec{r}_{01}{}^{2}}\right)  \ln\left(  \frac{\vec{r}_{14}{}^{2}%
}{\vec{r}_{04}{}^{2}}\right) \nonumber\\
+  &  \left.  \left(  \frac{\vec{r}_{13}{}^{2}}{\vec{r}_{01}{}^{2}\vec{r}%
_{03}{}^{2}}-\frac{\vec{r}_{14}{}^{2}}{\vec{r}_{01}{}^{2}\vec{r}_{04}{}^{2}%
}-\frac{\vec{r}_{23}{}^{2}}{\vec{r}_{02}{}^{2}\vec{r}_{03}{}^{2}}+\frac
{\vec{r}_{24}{}^{2}}{\vec{r}_{02}{}^{2}\vec{r}_{04}{}^{2}}\right)  \ln\left(
\frac{\vec{r}_{12}{}^{2}}{\vec{r}_{01}{}^{2}}\right)  \ln\left(  \frac{\vec
{r}_{12}{}^{2}}{\vec{r}_{02}{}^{2}}\right)  \right]  +(1\leftrightarrow
3,2\leftrightarrow4).
\end{align}%
\begin{align*}
\mathbf{\tilde{G}}_{1}  &  \mathbf{=}\frac{1}{2}(\mathbf{U}_{2^{\dag}%
0}\mathbf{U}_{0^{\dag}14^{\dag}3}+\mathbf{U}_{2^{\dag}0}\mathbf{U}_{0^{\dag
}34^{\dag}1}-\mathbf{U}_{0^{\dag}1}\mathbf{U}_{2^{\dag}04^{\dag}3}%
-\mathbf{U}_{0^{\dag}1}\mathbf{U}_{2^{\dag}34^{\dag}0})\left[  \ln\left(
\frac{\vec{r}_{23}{}^{2}}{\vec{r}_{02}{}^{2}}\right)  \ln\left(  \frac{\vec
{r}_{23}{}^{2}}{\vec{r}_{03}{}^{2}}\right)  \right. \\
&  \times\left(  \frac{\vec{r}_{13}{}^{2}}{\vec{r}_{01}{}^{2}\vec{r}_{03}%
{}^{2}}-\frac{\vec{r}_{12}{}^{2}}{\vec{r}_{01}{}^{2}\vec{r}_{02}{}^{2}}%
-\frac{\vec{r}_{23}{}^{2}}{\vec{r}_{02}{}^{2}\vec{r}_{03}{}^{2}}\right)
+\left(  \frac{\vec{r}_{23}{}^{2}}{\vec{r}_{02}{}^{2}\vec{r}_{03}{}^{2}}%
-\frac{\vec{r}_{12}{}^{2}}{\vec{r}_{01}{}^{2}\vec{r}_{02}{}^{2}}-\frac{\vec
{r}_{13}{}^{2}}{\vec{r}_{01}{}^{2}\vec{r}_{03}{}^{2}}\right)
\end{align*}%
\begin{align*}
&  \times\ln\left(  \frac{\vec{r}_{13}{}^{2}}{\vec{r}_{01}{}^{2}}\right)
\ln\left(  \frac{\vec{r}_{13}{}^{2}}{\vec{r}_{03}{}^{2}}\right)  +\ln\left(
\frac{\vec{r}_{14}{}^{2}}{\vec{r}_{01}{}^{2}}\right)  \ln\left(  \frac{\vec
{r}_{14}{}^{2}}{\vec{r}_{04}{}^{2}}\right)  \left(  \frac{\vec{r}_{12}{}^{2}%
}{\vec{r}_{01}{}^{2}\vec{r}_{02}{}^{2}}+\frac{\vec{r}_{14}{}^{2}}{\vec{r}%
_{01}{}^{2}\vec{r}_{04}{}^{2}}-\frac{\vec{r}_{24}{}^{2}}{\vec{r}_{02}{}%
^{2}\vec{r}_{04}{}^{2}}\right) \\
&  +\left.  \left(  \frac{\vec{r}_{12}{}^{2}}{\vec{r}_{01}{}^{2}\vec{r}_{02}%
{}^{2}}-\frac{\vec{r}_{14}{}^{2}}{\vec{r}_{01}{}^{2}\vec{r}_{04}{}^{2}}%
+\frac{\vec{r}_{24}{}^{2}}{\vec{r}_{02}{}^{2}\vec{r}_{04}{}^{2}}\right)
\ln\left(  \frac{\vec{r}_{24}{}^{2}}{\vec{r}_{02}{}^{2}}\right)  \ln\left(
\frac{\vec{r}_{24}{}^{2}}{\vec{r}_{04}{}^{2}}\right)  \right]
\end{align*}%
\begin{align}
+  &  \frac{1}{2}(2\mathbf{U}_{2^{\dag}1}\mathbf{U}_{4^{\dag}3}-N_{c}%
\mathbf{U}_{0^{\dag}12^{\dag}04^{\dag}3}-N_{c}\mathbf{U}_{0^{\dag}34^{\dag
}02^{\dag}1})\ln\left(  \frac{\vec{r}_{12}{}^{2}}{\vec{r}_{01}{}^{2}}\right)
\ln\left(  \frac{\vec{r}_{12}{}^{2}}{\vec{r}_{02}{}^{2}}\right) \nonumber\\
\times &  \left(  \frac{\vec{r}_{14}{}^{2}}{\vec{r}_{01}{}^{2}\vec{r}_{04}%
{}^{2}}-\frac{\vec{r}_{13}{}^{2}}{\vec{r}_{01}{}^{2}\vec{r}_{03}{}^{2}}%
+\frac{\vec{r}_{23}{}^{2}}{\vec{r}_{02}{}^{2}\vec{r}_{03}{}^{2}}-\frac{\vec
{r}_{24}{}^{2}}{\vec{r}_{02}{}^{2}\vec{r}_{04}{}^{2}}\right)
+(1\leftrightarrow3,2\leftrightarrow4).
\end{align}
As for the quadrupole, it is straightforward to check that none of the
functions $\mathbf{\tilde{G}}_{s},$ $\mathbf{\tilde{G}}_{a},$ $\mathbf{\tilde
{G}}_{\beta}$, $\mathbf{\tilde{G}}$ has unintegrable singularities.

\section{Quasi-conformal evolution equation for composite operators}

To construct composite conformal operators we use the prescription
\cite{Balitsky:2009xg} (see also \cite{NLOJIMWLKonformal})%
\begin{equation}
O^{conf}=O+\frac{1}{2}\frac{\partial O}{\partial\eta}\left\vert _{\frac
{\vec{r}_{mn}^{\,\,2}}{\vec{r}_{im}^{\,\,2}\vec{r}_{in}^{\,\,2}}%
\rightarrow\frac{\vec{r}_{mn}^{\,\,2}}{\vec{r}_{im}^{\,\,2}\vec{r}%
_{in}^{\,\,2}}\ln\left(  \frac{\vec{r}_{mn}^{\,\,2}a}{\vec{r}_{im}^{\,\,2}%
\vec{r}_{in}^{\,\,2}}\right)  }\right.  , \label{prescription}%
\end{equation}
where $a$ is an arbitrary constant. The conformal dipole reads
\cite{Balitsky:2009xg}
\begin{equation}
\mathbf{U}_{12^{\dag}}^{conf}=\mathbf{U}_{2^{\dag}1}+{\frac{\alpha_{s}}%
{4\pi^{2}}}\!\int\!d\vec{r}_{0}\frac{\vec{r}_{12}{}^{2}}{\vec{r}_{10}{}%
^{2}\vec{r}_{20}{}^{2}}\ln\left(  \frac{a\vec{r}_{12}{}^{2}}{\vec{r}_{10}%
{}^{2}\vec{r}_{20}{}^{2}}\right)  \left(  \mathbf{U}_{2^{\dag}0}%
\mathbf{U}_{0^{\dag}1}-N_{c}\mathbf{U}_{2^{\dag}1}\right)  .
\end{equation}
The evolution equation for this operator \cite{Balitsky:2009xg} is
quasi-conformal%
\begin{align}
\frac{\partial\mathbf{U}_{12^{\dag}}^{conf}}{\partial\eta}=  &  \frac
{\alpha_{s}}{2\pi^{2}}\int\!d\vec{r}_{0}\frac{\vec{r}_{12}{}^{2}}{\vec{r}%
_{10}{}^{2}\vec{r}_{20}{}^{2}}\left(  1+\frac{\alpha_{s}}{2\pi}M_{12}^{\beta
}\right)  \left(  \mathbf{U}_{2^{\dag}0}\mathbf{U}_{0^{\dag}1}-N_{c}%
\mathbf{U}_{2^{\dag}1}\right)  ^{conf}\nonumber\\
+  &  {\frac{\alpha_{s}^{2}}{4\pi^{4}}}\!\int\!d\vec{r}_{0}d\vec{r}%
_{5}\{L_{12}^{C}((\mathbf{U}_{0^{\dag}52^{\dag}05^{\dag}1}-\mathbf{U}%
_{0^{\dag}1}\mathbf{U}_{2^{\dag}5}\mathbf{U}_{5^{\dag}0}-(0\rightarrow
5))+(0\leftrightarrow5))\nonumber\\
+  &  \tilde{L}_{12}^{C}(\mathbf{U}_{0^{\dag}5}\mathbf{U}_{2^{\dag}%
0}\mathbf{U}_{5^{\dag}1}-(0\leftrightarrow5))\nonumber\\
-  &  2n_{f}L_{12}^{q}(tr(t^{a}U_{1}t^{b}U_{2}^{\dag})tr(t^{a}U_{5}t^{b}%
(U_{0}^{\dag}-U_{5}^{\dag}))+(5\leftrightarrow0))\}, \label{NLOBKC}%
\end{align}
where $M_{12}^{\beta}$ is defined in (\ref{Mbeta}); $L_{ij}^{C}\equiv
L^{C}(\vec{r}_{i},\vec{r}_{j})$ and $\tilde{L}_{ij}^{C}\equiv\tilde{L}%
^{C}(\vec{r}_{i},\vec{r}_{j})$ were introduced in this form in \cite{bg}%
\begin{equation}
L_{12}^{C}=L_{12}+\frac{\vec{r}_{12}{}^{2}}{4\vec{r}_{01}{}^{2}\vec{r}_{05}%
{}^{2}\vec{r}_{25}{}^{2}}\ln\left(  \frac{\vec{r}_{02}{}^{2}\vec{r}_{15}{}%
^{2}}{\vec{r}_{05}{}^{2}\vec{r}_{12}{}^{2}}\right)  +\frac{\vec{r}_{12}{}^{2}%
}{4\vec{r}_{02}{}^{2}\vec{r}_{05}{}^{2}\vec{r}_{15}{}^{2}}\ln\left(
\frac{\vec{r}_{01}{}^{2}\vec{r}_{25}{}^{2}}{\vec{r}_{05}{}^{2}\vec{r}_{12}%
{}^{2}}\right)  , \label{LC}%
\end{equation}%
\begin{equation}
\tilde{L}_{12}^{C}=\tilde{L}_{12}+\frac{\vec{r}_{12}{}^{2}}{4\vec{r}_{01}%
{}^{2}\vec{r}_{05}{}^{2}\vec{r}_{25}{}^{2}}\ln\left(  \frac{\vec{r}_{02}{}%
^{2}\vec{r}_{15}{}^{2}}{\vec{r}_{05}{}^{2}\vec{r}_{12}{}^{2}}\right)
-\frac{\vec{r}_{12}{}^{2}}{4\vec{r}_{02}{}^{2}\vec{r}_{05}{}^{2}\vec{r}_{15}%
{}^{2}}\ln\left(  \frac{\vec{r}_{01}{}^{2}\vec{r}_{25}{}^{2}}{\vec{r}_{05}%
{}^{2}\vec{r}_{12}{}^{2}}\right)  . \label{LCtilde}%
\end{equation}
For the conformal quadrupole operator using (\ref{LOkernel}) we have%
\begin{align}
\mathbf{U}_{12^{\dag}34^{\dag}}^{conf}=  &  \mathbf{U}_{12^{\dag}34^{\dag}%
}+{\frac{\alpha_{s}}{8\pi^{2}}}\!\int\!d\vec{r}_{0}\nonumber\\
\times &  \left\{  \frac{\vec{r}_{14}{}^{2}}{\vec{r}_{10}{}^{2}\vec{r}_{40}%
{}^{2}}\ln\left(  \frac{a\vec{r}_{14}{}^{2}}{\vec{r}_{10}{}^{2}\vec{r}_{40}%
{}^{2}}\right)  \left(  \mathbf{U}_{10^{\dag}}\mathbf{U}_{02^{\dag}34^{\dag}%
}+\mathbf{U}_{4^{\dag}0}\mathbf{U}_{12^{\dag}30^{\dag}}-\left(  0\rightarrow
1\right)  \right)  \right. \nonumber\\
+  &  \frac{\vec{r}_{12}{}^{2}}{\vec{r}_{10}{}^{2}\vec{r}_{20}{}^{2}}%
\ln\left(  \frac{a\vec{r}_{12}{}^{2}}{\vec{r}_{10}{}^{2}\vec{r}_{20}{}^{2}%
}\right)  \left(  \mathbf{U}_{10^{\dag}}\mathbf{U}_{02^{\dag}34^{\dag}%
}+\mathbf{U}_{2^{\dag}0}\mathbf{U}_{10^{\dag}34^{\dag}}-\left(  0\rightarrow
1\right)  \right) \nonumber\\
-  &  \frac{\vec{r}_{24}{}^{2}}{2\vec{r}_{20}{}^{2}\vec{r}_{40}{}^{2}}%
\ln\left(  \frac{a\vec{r}_{24}{}^{2}}{\vec{r}_{20}{}^{2}\vec{r}_{40}{}^{2}%
}\right)  \left(  \mathbf{U}_{10^{\dag}}\mathbf{U}_{02^{\dag}34^{\dag}%
}+\mathbf{U}_{30^{\dag}}\mathbf{U}_{04^{\dag}12^{\dag}}-(0\rightarrow4)\right)
\nonumber\\
-  &  \frac{\vec{r}_{13}{}^{2}}{2\vec{r}_{10}{}^{2}\vec{r}_{30}{}^{2}}%
\ln\left(  \frac{a\vec{r}_{13}{}^{2}}{\vec{r}_{10}{}^{2}\vec{r}_{30}{}^{2}%
}\right)  \left(  \mathbf{U}_{4^{\dag}0}\mathbf{U}_{12^{\dag}30^{\dag}%
}+\mathbf{U}_{2^{\dag}0}\mathbf{U}_{34^{\dag}10^{\dag}}-(0\rightarrow1)\right)
\nonumber\\
+  &  \left.  (1\leftrightarrow3,2\leftrightarrow4)\right\}  . \label{qc}%
\end{align}
The conformal double dipole operator reads%
\begin{align}
(\mathbf{U}_{12^{\dag}}\mathbf{U}_{34^{\dag}})^{conf}=  &  \mathbf{U}%
_{12^{\dag}}\mathbf{U}_{34^{\dag}}+{\frac{\alpha_{s}}{8\pi^{2}}}\!\int
\!d\vec{r}_{0}(\mathbf{U}_{2^{\dag}14^{\dag}3}+\mathbf{U}_{2^{\dag}34^{\dag}%
1}-\mathbf{U}_{2^{\dag}10^{\dag}34^{\dag}0}-\mathbf{U}_{2^{\dag}04^{\dag
}30^{\dag}1})\nonumber\\
\times &  \left(  \frac{\vec{r}_{13}{}^{2}}{\vec{r}_{10}{}^{2}\vec{r}_{30}%
{}^{2}}\ln\left(  \frac{a\vec{r}_{13}{}^{2}}{\vec{r}_{10}{}^{2}\vec{r}_{30}%
{}^{2}}\right)  -\frac{\vec{r}_{23}{}^{2}}{\vec{r}_{20}{}^{2}\vec{r}_{30}%
{}^{2}}\ln\left(  \frac{a\vec{r}_{23}{}^{2}}{\vec{r}_{20}{}^{2}\vec{r}_{30}%
{}^{2}}\right)  \right. \nonumber\\
-  &  \left.  \frac{\vec{r}_{14}{}^{2}}{\vec{r}_{10}{}^{2}\vec{r}_{40}{}^{2}%
}\ln\left(  \frac{a\vec{r}_{14}{}^{2}}{\vec{r}_{10}{}^{2}\vec{r}_{40}{}^{2}%
}\right)  +\frac{\vec{r}_{24}{}^{2}}{\vec{r}_{20}{}^{2}\vec{r}_{40}{}^{2}}%
\ln\left(  \frac{a\vec{r}_{24}{}^{2}}{\vec{r}_{20}{}^{2}\vec{r}_{40}{}^{2}%
}\right)  \right) \nonumber\\
+  &  \mathbf{U}_{4^{\dag}3}(\mathbf{U}_{2^{\dag}1}^{conf}\mathbf{-U}%
_{2^{\dag}1})+\mathbf{U}_{2^{\dag}1}(\mathbf{U}_{4^{\dag}3}^{conf}%
\mathbf{-U}_{4^{\dag}3}). \label{ddc}%
\end{align}
The evolution equations for the conformal quadrupole and double dipole
operators in the conformal basis have the general form%
\begin{align}
\frac{\partial\mathbf{U}_{12^{\dag}34^{\dag}}^{conf}}{\partial\eta}=  &
{\frac{\alpha_{s}}{4\pi^{2}}}\!\int\!d\vec{r}_{0}\left\{  \frac{\vec{r}_{14}%
{}^{2}}{\vec{r}_{10}{}^{2}\vec{r}_{40}{}^{2}}\left(  \mathbf{U}_{10^{\dag}%
}\mathbf{U}_{02^{\dag}34^{\dag}}+\mathbf{U}_{4^{\dag}0}\mathbf{U}_{12^{\dag
}30^{\dag}}-\left(  0\rightarrow1\right)  \right)  ^{conf}\right. \nonumber\\
+  &  \frac{\vec{r}_{12}{}^{2}}{\vec{r}_{10}{}^{2}\vec{r}_{20}{}^{2}}\left(
\mathbf{U}_{10^{\dag}}\mathbf{U}_{02^{\dag}34^{\dag}}+\mathbf{U}_{2^{\dag}%
0}\mathbf{U}_{10^{\dag}34^{\dag}}-\left(  0\rightarrow1\right)  \right)
^{conf}\nonumber\\
-  &  \frac{\vec{r}_{24}{}^{2}}{2\vec{r}_{20}{}^{2}\vec{r}_{40}{}^{2}}\left(
\mathbf{U}_{10^{\dag}}\mathbf{U}_{02^{\dag}34^{\dag}}+\mathbf{U}_{30^{\dag}%
}\mathbf{U}_{04^{\dag}12^{\dag}}-(0\rightarrow4)\right)  ^{conf}\nonumber\\
-  &  \left.  \frac{\vec{r}_{13}{}^{2}}{2\vec{r}_{10}{}^{2}\vec{r}_{30}{}^{2}%
}\left(  \mathbf{U}_{4^{\dag}0}\mathbf{U}_{12^{\dag}30^{\dag}}+\mathbf{U}%
_{2^{\dag}0}\mathbf{U}_{34^{\dag}10^{\dag}}-(0\rightarrow1)\right)
^{conf}+(1\leftrightarrow3,2\leftrightarrow4)\right\} \nonumber\\
+  &  {\frac{\alpha_{s}^{2}}{8\pi^{4}}}\!\int\!d\vec{r}_{0}d\vec{r}%
_{5}~\mathbf{(\mathbf{G}}_{s}^{conf}\mathbf{\mathbf{+G}}_{a}^{conf}%
\mathbf{)+}{\frac{\alpha_{s}^{2}}{8\pi^{3}}}\!\int\!d\vec{r}_{0}%
~(\mathbf{G}_{\beta}+\mathbf{G}^{conf}\mathbf{)}, \label{GdefinitionC}%
\end{align}%
\begin{align}
\frac{\partial(\mathbf{U}_{12^{\dag}}\mathbf{U}_{34^{\dag}})^{conf}}%
{\partial\eta}  &  =\left\{  {\frac{\alpha_{s}}{8\pi^{2}}}\!\int\!d\vec{r}%
_{0}\left[  \frac{4\vec{r}_{12}{}^{2}}{\vec{r}_{10}{}^{2}\vec{r}_{20}{}^{2}%
}\left(  \mathbf{U}_{4^{\dag}3}\mathbf{U}_{2^{\dag}0}\mathbf{U}_{0^{\dag}%
1}-N_{c}\mathbf{U}_{4^{\dag}3}\mathbf{U}_{2^{\dag}1}\right)  ^{conf}\right.
\right. \nonumber\\
&  +\left(  \frac{\vec{r}_{13}{}^{2}}{\vec{r}_{10}{}^{2}\vec{r}_{30}{}^{2}%
}-\frac{\vec{r}_{23}{}^{2}}{\vec{r}_{20}{}^{2}\vec{r}_{30}{}^{2}}-\frac
{\vec{r}_{14}{}^{2}}{\vec{r}_{10}{}^{2}\vec{r}_{40}{}^{2}}+\frac{\vec{r}%
_{24}{}^{2}}{\vec{r}_{20}{}^{2}\vec{r}_{40}{}^{2}}\right) \nonumber\\
&  \times\left.  (\mathbf{U}_{2^{\dag}14^{\dag}3}+\mathbf{U}_{2^{\dag}%
34^{\dag}1}-\mathbf{U}_{2^{\dag}10^{\dag}34^{\dag}0}-\mathbf{U}_{2^{\dag
}04^{\dag}30^{\dag}1})^{conf}\right] \nonumber\\
&  +\left.  \mathbf{U}_{34^{\dag}}\langle K_{NLO}\otimes\mathbf{U}_{12^{\dag}%
}^{conf}\rangle+(1\leftrightarrow3,2\leftrightarrow4)\right\} \nonumber\\
&  +{\frac{\alpha_{s}^{2}}{8\pi^{4}}}\!\int\!d\vec{r}_{0}d\vec{r}%
_{5}~\mathbf{(\mathbf{\tilde{G}}}_{s}^{conf}\mathbf{\mathbf{+\tilde{G}}}%
_{a}^{conf}\mathbf{)+}{\frac{\alpha_{s}^{2}}{8\pi^{3}}}\!\int\!d\vec{r}%
_{0}~(\mathbf{\tilde{G}}_{\beta}+\mathbf{\tilde{G}}^{conf}\mathbf{)}.
\label{GtildedefinitionC}%
\end{align}
As in the previous section, the individual NLO evolution of the dipoles here
is taken out of the functions $\mathbf{\tilde{G}}$%
\begin{equation}
\langle K_{NLO}\otimes\mathbf{U}_{12^{\dag}}^{conf}\rangle=\frac
{\partial\mathbf{U}_{12^{\dag}}^{conf}}{\partial\eta}-\frac{\alpha_{s}}%
{2\pi^{2}}\int\!d\vec{r}_{0}\frac{\vec{r}_{12}{}^{2}}{\vec{r}_{10}{}^{2}%
\vec{r}_{20}{}^{2}}\left(  \mathbf{U}_{2^{\dag}0}\mathbf{U}_{0^{\dag}1}%
-N_{c}\mathbf{U}_{2^{\dag}1}\right)  ^{conf}.
\end{equation}
Therefore one can rewrite (\ref{GtildedefinitionC})
\begin{align}
\frac{\partial(\mathbf{U}_{12^{\dag}}\mathbf{U}_{34^{\dag}})^{conf}}%
{\partial\eta}=  &  \left\{  \mathbf{U}_{34^{\dag}}\frac{\partial
\mathbf{U}_{12^{\dag}}^{conf}}{\partial\eta}+{\frac{\alpha_{s}}{8\pi^{2}}%
}\!\int\!d\vec{r}_{0}\left[  \frac{4\vec{r}_{12}{}^{2}}{\vec{r}_{10}{}^{2}%
\vec{r}_{20}{}^{2}}\right.  \right. \nonumber\\
\times &  \{\left(  \mathbf{U}_{4^{\dag}3}\mathbf{U}_{2^{\dag}0}%
\mathbf{U}_{0^{\dag}1}-N_{c}\mathbf{U}_{4^{\dag}3}\mathbf{U}_{2^{\dag}%
1}\right)  ^{conf}-\mathbf{U}_{4^{\dag}3}\left(  \mathbf{U}_{2^{\dag}%
0}\mathbf{U}_{0^{\dag}1}-N_{c}\mathbf{U}_{2^{\dag}1}\right)  ^{conf}%
\}\nonumber\\
+  &  \left(  \frac{\vec{r}_{13}{}^{2}}{\vec{r}_{10}{}^{2}\vec{r}_{30}{}^{2}%
}-\frac{\vec{r}_{23}{}^{2}}{\vec{r}_{20}{}^{2}\vec{r}_{30}{}^{2}}-\frac
{\vec{r}_{14}{}^{2}}{\vec{r}_{10}{}^{2}\vec{r}_{40}{}^{2}}+\frac{\vec{r}%
_{24}{}^{2}}{\vec{r}_{20}{}^{2}\vec{r}_{40}{}^{2}}\right) \nonumber\\
\times &  \left.  \left.  (\mathbf{U}_{2^{\dag}14^{\dag}3}+\mathbf{U}%
_{2^{\dag}34^{\dag}1}-\mathbf{U}_{2^{\dag}10^{\dag}34^{\dag}0}-\mathbf{U}%
_{2^{\dag}04^{\dag}30^{\dag}1})^{conf}\right]  +(1\leftrightarrow
3,2\leftrightarrow4)\right\} \nonumber\\
+  &  {\frac{\alpha_{s}^{2}}{8\pi^{4}}}\!\int\!d\vec{r}_{0}d\vec{r}%
_{5}~\mathbf{(\mathbf{\tilde{G}}}_{s}^{conf}\mathbf{\mathbf{+\tilde{G}}}%
_{a}^{conf}\mathbf{)+}{\frac{\alpha_{s}^{2}}{8\pi^{3}}}\!\int\!d\vec{r}%
_{0}~(\mathbf{\tilde{G}}_{\beta}+\mathbf{\tilde{G}}^{conf}\mathbf{)}.
\end{align}
Plainly, $\mathbf{G}_{\beta}$ and $\mathbf{\tilde{G}}_{\beta}$ are the same as
in (\ref{Gdefinition}) and (\ref{Gtildedefinition}). The other functions
$\mathbf{G}^{conf}$ will be given below.

To obtain these functions one has to calculate the evolution equations for
conformal operators (\ref{qc}, \ref{ddc})\ using (\ref{LOdipole}--\ref{sex})
and express the results in terms of conformal operators via
(\ref{prescription}). Technically, it means that one has to add to the kernels
of the evolution equations from the previous section the corrections in the
form of double integrals w.r.t. $\vec{r}_{0}$ and $\vec{r}_{5}$
\cite{Balitsky:2009xg}. To get the conformally invariant results one has to
symmetrize these corrections according to (\ref{step1}). Then, the terms with
color operators independent of $\vec{r}_{0}$ (or $\vec{r}_{5}$) can be
integrated w.r.t. $\vec{r}_{0}$ (or $\vec{r}_{5}$) via the integrals from
appendix A of \cite{Fadin:2009za}. Finally, the terms with color operators
independent of both $\vec{r}_{0}$ and $\vec{r}_{5}$ can be integrated with
respect to both $\vec{r}_{0}$ and $\vec{r}_{5}.$ In addition to the integrals
from appendix A of \cite{Fadin:2009za}, one needs the following integral
\begin{equation}
\int d\vec{r}_{0}\left(  \frac{\vec{r}_{01}\vec{r}_{02}}{\vec{r}_{01}%
^{\,\,2}\vec{r}_{02}^{\,\,2}}-\frac{\vec{r}_{01}\vec{r}_{03}}{\vec{r}%
_{01}^{\,\,2}\vec{r}_{03}^{\,\,2}}\right)  \ln^{2}\left(  \frac{\vec{r}%
_{02}^{\,\,2}\vec{r}_{13}^{\,\,2}}{\vec{r}_{03}^{\,\,2}\vec{r}_{12}^{\,\,2}%
}\right)  =\frac{\pi}{3}\ln^{3}\left(  \frac{\vec{r}_{13}^{\,\,2}}{\vec
{r}_{12}^{\,\,2}}\right)  .
\end{equation}

\subsection{Quadrupole}

For the symmetric contribution $\mathbf{\mathbf{G}}_{s}^{conf}$ we have%
\begin{equation}
\mathbf{\mathbf{G}}_{s}^{conf}=\mathbf{\mathbf{G}}_{s1}^{conf}+n_{f}%
\mathbf{G}_{q}+\mathbf{\mathbf{G}}_{s2}^{conf},
\end{equation}
where $\mathbf{G}_{q}$ did not change. It is defined in (\ref{Gq}) .
\begin{align}
\mathbf{\mathbf{G}}_{s1}^{conf}=  &  \left(  \left\{  \mathbf{U}_{0^{\dag
}34^{\dag}15^{\dag}02^{\dag}5}-\mathbf{U}_{5^{\dag}0}\mathbf{U}_{2^{\dag}%
5}\mathbf{U}_{0^{\dag}34^{\dag}1}-(5\rightarrow0)\right\}  +(5\leftrightarrow
0)\right)  \left(  L_{12}^{C}+L_{32}^{C}-L_{13}^{C}\right) \nonumber\\
+  &  \left(  \left\{  \mathbf{U}_{0^{\dag}15^{\dag}02^{\dag}34^{\dag}%
5}-\mathbf{U}_{0^{\dag}5}\mathbf{U}_{5^{\dag}1}\mathbf{U}_{2^{\dag}34^{\dag}%
0}-(5\rightarrow0)\right\}  +(5\leftrightarrow0)\right)  (L_{12}^{C}%
+L_{14}^{C}-L_{42}^{C})\nonumber\\
+  &  (1\leftrightarrow3,2\leftrightarrow4),
\end{align}
where $L^{C}$ is defined in (\ref{LC}).%
\begin{align}
\mathbf{\mathbf{G}}_{s2}^{conf}\mathbf{=}  &  \frac{1}{2}\left(
\mathbf{U}_{0^{\dag}34^{\dag}52^{\dag}05^{\dag}1}-\mathbf{U}_{0^{\dag}%
1}\mathbf{U}_{2^{\dag}5}\mathbf{U}_{4^{\dag}05^{\dag}3}+(5\leftrightarrow
0)\right) \nonumber\\
\times &  (M_{1}^{C34}-M_{1}^{C43}+M_{1}^{C42}-M_{1}^{C24}+M_{2}^{C43}%
-M_{2}^{C34}+M_{2}^{C31}-M_{2}^{C13})\nonumber\\
+  &  \frac{1}{2}\left(  \mathbf{U}_{0^{\dag}35^{\dag}02^{\dag}54^{\dag}%
1}-\mathbf{U}_{0^{\dag}3}\mathbf{U}_{2^{\dag}5}\mathbf{U}_{4^{\dag}15^{\dag}%
0}+(5\leftrightarrow0)\right) \nonumber\\
\times &  (M_{3}^{C14}-M_{3}^{C41}+M_{3}^{C42}-M_{3}^{C24}+M_{2}^{C41}%
-M_{2}^{C14}+M_{2}^{C13}-M_{2}^{C31})\nonumber\\
+  &  \frac{1}{2}\left(  \mathbf{U}_{0^{\dag}15^{\dag}02^{\dag}34^{\dag}%
5}-\mathbf{U}_{0^{\dag}5}\mathbf{U}_{5^{\dag}1}\mathbf{U}_{2^{\dag}34^{\dag}%
0}+(5\leftrightarrow0)\right)  (M_{2}^{C14}-M_{2}^{C41}+M_{4}^{C12}%
-M_{4}^{C21})\nonumber\\
+  &  \frac{1}{2}\left(  \mathbf{U}_{0^{\dag}34^{\dag}15^{\dag}02^{\dag}%
5}-\mathbf{U}_{5^{\dag}0}\mathbf{U}_{2^{\dag}5}\mathbf{U}_{0^{\dag}34^{\dag}%
1}+(5\leftrightarrow0)\right)  (M_{1}^{C23}-M_{1}^{C32}+M_{3}^{C21}%
-M_{3}^{C12})\nonumber\\
+  &  (1\leftrightarrow3,2\leftrightarrow4).
\end{align}
Here $M_{i}^{Cjk}\equiv M^{C}(\vec{r}_{i},\vec{r}_{j},\vec{r}_{k})$ reads%
\begin{align}
M_{2}^{C13}=  &  M_{2}^{13}+\frac{\vec{r}_{15}{}^{2}\vec{r}_{23}{}^{2}}%
{8\vec{r}_{01}{}^{2}\vec{r}_{05}{}^{2}\vec{r}_{25}{}^{2}\vec{r}_{35}{}^{2}}%
\ln\left(  \frac{\vec{r}_{01}{}^{2}\vec{r}_{05}{}^{2}\vec{r}_{23}{}^{2}}%
{\vec{r}_{15}{}^{2}\vec{r}_{25}{}^{2}\vec{r}_{35}{}^{2}}\right)  -\frac
{\vec{r}_{12}{}^{2}}{4\vec{r}_{02}{}^{2}\vec{r}_{05}{}^{2}\vec{r}_{15}{}^{2}%
}\ln\left(  \frac{\vec{r}_{01}{}^{2}\vec{r}_{25}{}^{2}}{\vec{r}_{05}{}^{2}%
\vec{r}_{12}{}^{2}}\right) \nonumber\\
+  &  \frac{\vec{r}_{13}{}^{2}}{8\vec{r}_{01}{}^{2}\vec{r}_{05}{}^{2}\vec
{r}_{35}{}^{2}}\ln\left(  \frac{\vec{r}_{05}{}^{2}\vec{r}_{13}{}^{2}\vec
{r}_{35}{}^{2}}{\vec{r}_{01}{}^{2}\vec{r}_{03}{}^{4}{}}\right)  -\frac{\vec
{r}_{12}{}^{2}\vec{r}_{23}{}^{2}}{8\vec{r}_{01}{}^{2}\vec{r}_{02}{}^{2}\vec
{r}_{25}{}^{2}\vec{r}_{35}{}^{2}}\ln\left(  \frac{\vec{r}_{01}{}^{2}\vec
{r}_{02}{}^{2}\vec{r}_{23}{}^{2}}{\vec{r}_{12}{}^{2}\vec{r}_{25}{}^{2}\vec
{r}_{35}{}^{2}}\right) \nonumber\\
-  &  \frac{\vec{r}_{03}{}^{2}\vec{r}_{12}{}^{2}}{8\vec{r}_{01}{}^{2}\vec
{r}_{02}{}^{2}\vec{r}_{05}{}^{2}\vec{r}_{35}{}^{2}}\ln\left(  \frac{\vec
{r}_{05}{}^{2}\vec{r}_{12}{}^{2}\vec{r}_{35}{}^{2}}{\vec{r}_{01}{}^{2}\vec
{r}_{02}{}^{2}\vec{r}_{03}{}^{2}}\right) \nonumber\\
=  &  \frac{\vec{r}_{15}{}^{2}\vec{r}_{23}{}^{2}}{8\vec{r}_{01}{}^{2}\vec
{r}_{05}{}^{2}\vec{r}_{25}{}^{2}\vec{r}_{35}{}^{2}}\ln\left(  \frac{\vec
{r}_{01}{}^{2}\vec{r}_{05}{}^{2}\vec{r}_{23}{}^{2}\vec{r}_{25}{}^{2}}{\vec
{r}_{02}{}^{4}{}\vec{r}_{15}{}^{2}\vec{r}_{35}{}^{2}}\right)  -\frac{\vec
{r}_{12}{}^{2}}{4\vec{r}_{02}{}^{2}\vec{r}_{05}{}^{2}\vec{r}_{15}{}^{2}}%
\ln\left(  \frac{\vec{r}_{01}{}^{2}\vec{r}_{25}{}^{2}}{\vec{r}_{05}{}^{2}%
\vec{r}_{12}{}^{2}}\right) \nonumber\\
-  &  \frac{\vec{r}_{13}{}^{2}}{8\vec{r}_{01}{}^{2}\vec{r}_{05}{}^{2}\vec
{r}_{35}{}^{2}}\ln\left(  \frac{\vec{r}_{01}{}^{2}\vec{r}_{03}{}^{4}{}\vec
{r}_{25}{}^{4}{}}{\vec{r}_{02}{}^{4}{}\vec{r}_{05}{}^{2}\vec{r}_{13}{}^{2}%
\vec{r}_{35}{}^{2}}\right)  +\frac{\vec{r}_{23}{}^{2}\vec{r}_{12}{}^{2}}%
{8\vec{r}_{01}{}^{2}\vec{r}_{02}{}^{2}\vec{r}_{25}{}^{2}\vec{r}_{35}{}^{2}}%
\ln\left(  \frac{\vec{r}_{02}{}^{2}\vec{r}_{12}{}^{2}\vec{r}_{35}{}^{2}}%
{\vec{r}_{01}{}^{2}\vec{r}_{23}{}^{2}\vec{r}_{25}{}^{2}}\right) \nonumber\\
+  &  \frac{\vec{r}_{03}{}^{2}\vec{r}_{12}{}^{2}}{8\vec{r}_{01}{}^{2}\vec
{r}_{02}{}^{2}\vec{r}_{05}{}^{2}\vec{r}_{35}{}^{2}}\ln\left(  \frac{\vec
{r}_{01}{}^{2}\vec{r}_{03}{}^{2}\vec{r}_{25}{}^{4}{}}{\vec{r}_{02}{}^{2}%
\vec{r}_{05}{}^{2}\vec{r}_{12}{}^{2}\vec{r}_{35}{}^{2}}\right)  .
\end{align}
The function $M_{i}^{Cjk}$ is conformally invariant. It does not have property
(\ref{Masimproperty}). Nevertheless like $\mathbf{\mathbf{G}}_{s2},$
$\mathbf{\mathbf{G}}_{s2}^{conf}$ can be rearranged to form
(\ref{Gs2vanishproof}) where instead of $M_{i}^{jk}$ will be $M_{i}^{Cjk}.$ As
a result $\mathbf{\mathbf{G}}_{s2}^{conf}$ vanishes without the shockwave.
Finally, one can see that $\mathbf{\mathbf{G}}_{s}^{conf}$ can be formally
obtained from $\mathbf{\mathbf{G}}_{s}$ via the substitution $M\rightarrow
M^{C},L\rightarrow L^{C}.$

The contribution which is the product of the antisymmetric w.r.t.
$5\leftrightarrow0$ parts reads
\begin{equation}
\mathbf{G}_{a}^{conf}=\mathbf{G}_{a1}^{conf}+\mathbf{G}_{a2}^{conf}%
+\mathbf{G}_{a3}^{conf}.
\end{equation}%
\begin{align}
\mathbf{G}_{a1}^{conf}\mathbf{=}  &  \frac{1}{2}\left(  \mathbf{U}_{0^{\dag}%
1}\mathbf{U}_{2^{\dag}5}\mathbf{U}_{4^{\dag}05^{\dag}3}+\mathbf{U}_{0^{\dag
}34^{\dag}52^{\dag}05^{\dag}1}-(5\leftrightarrow0)\right) \nonumber\\
\times &  (M_{2}^{C31}+M_{2}^{C13}-M_{2}^{C34}-M_{2}^{C43}-M_{1}^{C42}%
-M_{1}^{C24}+M_{1}^{C43}+M_{1}^{C34}-R_{2134})\nonumber\\
+  &  \frac{1}{2}\left(  \mathbf{U}_{0^{\dag}3}\mathbf{U}_{2^{\dag}%
5}\mathbf{U}_{4^{\dag}15^{\dag}0}+\mathbf{U}_{0^{\dag}35^{\dag}02^{\dag
}54^{\dag}1}-(5\leftrightarrow0)\right) \nonumber\\
\times &  (M_{2}^{C13}+M_{2}^{C31}-M_{2}^{C14}-M_{2}^{C41}-M_{3}^{C42}%
-M_{3}^{C24}+M_{3}^{C41}+M_{3}^{C14}+R_{3241})\nonumber\\
+  &  (1\leftrightarrow3,2\leftrightarrow4).
\end{align}
Here $R_{ijkl}\equiv R(\vec{r}_{i},\vec{r}_{j},\vec{r}_{k},\vec{r}_{l})$ is a
conformally invariant function. It reads%
\begin{align}
R_{2134}=  &  \frac{\vec{r}_{12}{}^{2}}{2\vec{r}_{01}{}^{2}\vec{r}_{05}{}%
^{2}\vec{r}_{25}{}^{2}}\ln\left(  \frac{\vec{r}_{02}{}^{2}\vec{r}_{15}{}^{2}%
}{\vec{r}_{05}{}^{2}\vec{r}_{12}{}^{2}}\right)  -\frac{\vec{r}_{12}{}^{2}%
}{2\vec{r}_{02}{}^{2}\vec{r}_{05}{}^{2}\vec{r}_{15}{}^{2}}\ln\left(
\frac{\vec{r}_{01}{}^{2}\vec{r}_{25}{}^{2}}{\vec{r}_{05}{}^{2}\vec{r}_{12}%
{}^{2}}\right) \nonumber\\
+  &  \frac{\vec{r}_{24}{}^{2}}{2\vec{r}_{02}{}^{2}\vec{r}_{05}{}^{2}\vec
{r}_{45}{}^{2}}\ln\left(  \frac{\vec{r}_{04}{}^{2}\vec{r}_{25}{}^{2}}{\vec
{r}_{05}{}^{2}\vec{r}_{24}{}^{2}}\right)  -\frac{\vec{r}_{13}{}^{2}}{2\vec
{r}_{01}{}^{2}\vec{r}_{05}{}^{2}\vec{r}_{35}{}^{2}}\ln\left(  \frac{\vec
{r}_{03}{}^{2}\vec{r}_{15}{}^{2}}{\vec{r}_{05}{}^{2}\vec{r}_{13}{}^{2}%
}\right)  .
\end{align}%
\begin{align}
\mathbf{G}_{a2}^{conf}\mathbf{=}  &  \frac{1}{2}\left(  \mathbf{U}_{0^{\dag
}34^{\dag}15^{\dag}02^{\dag}5}-(5\leftrightarrow0)\right)  (R_{231}%
-R_{123}\nonumber\\
+  &  2\tilde{L}_{13}^{C}-M_{1}^{C23}-M_{1}^{C32}+M_{3}^{C21}+M_{3}%
^{C12}+M_{2}^{C11}+M_{1}^{C22}-M_{3}^{C22}-M_{2}^{C33})\nonumber\\
+  &  \frac{1}{2}\left(  \mathbf{U}_{0^{\dag}15^{\dag}02^{\dag}34^{\dag}%
5}-(5\leftrightarrow0)\right)  (R_{124}-R_{142}\nonumber\\
+  &  2\tilde{L}_{42}^{C}-M_{1}^{C22}-M_{2}^{C11}+M_{1}^{C44}+M_{4}%
^{C11}+M_{2}^{C14}+M_{2}^{C41}-M_{4}^{C21}-M_{4}^{C12})\nonumber\\
+  &  (1\leftrightarrow3,2\leftrightarrow4).
\end{align}
Here the function $\tilde{L}^{C}$ is defined in (\ref{LCtilde}) and
$R_{ijk}\equiv R(\vec{r}_{i},\vec{r}_{j},\vec{r}_{k})$ is another conformally
invariant function. It reads%
\begin{equation}
R_{123}=\frac{\vec{r}_{13}{}^{2}}{2\vec{r}_{01}{}^{2}\vec{r}_{05}{}^{2}\vec
{r}_{35}{}^{2}}\ln\left(  \frac{\vec{r}_{03}{}^{2}\vec{r}_{15}{}^{2}}{\vec
{r}_{05}{}^{2}\vec{r}_{13}{}^{2}}\right)  +\frac{\vec{r}_{23}{}^{2}}{2\vec
{r}_{02}{}^{2}\vec{r}_{05}{}^{2}\vec{r}_{35}{}^{2}}\ln\left(  \frac{\vec
{r}_{03}{}^{2}\vec{r}_{25}{}^{2}}{\vec{r}_{05}{}^{2}\vec{r}_{23}{}^{2}%
}\right)  .
\end{equation}
In fact, there is freedom in the definition of the functions $M_{i}^{Cjk},$
$R_{ijk}$ and $R_{ijkl}$ since one can redistribute terms between them. For
example, one can try to redefine $M_{i}^{Cjk}$ so that to make the functions
$R$ zero.

The remaining antisymmetric contribution reads%
\begin{align}
\mathbf{G}_{a3}^{conf}\mathbf{=}  &  \frac{1}{2}\left(  \mathbf{U}_{0^{\dag}%
5}\mathbf{U}_{5^{\dag}1}\mathbf{U}_{2^{\dag}34^{\dag}0}-(5\leftrightarrow
0)\right) \nonumber\\
\times &  (2\tilde{L}_{12}^{C}+2\tilde{L}_{14}^{C}-M_{2}^{C44}-M_{4}%
^{C22}+M_{2}^{C14}+M_{2}^{C41}+M_{4}^{C21}+M_{4}^{C12}+R_{241})\nonumber\\
+  &  \frac{1}{2}\left(  \mathbf{U}_{5^{\dag}0}\mathbf{U}_{2^{\dag}%
5}\mathbf{U}_{0^{\dag}34^{\dag}1}-(5\leftrightarrow0)\right) \nonumber\\
\times &  (2\tilde{L}_{21}^{C}+2\tilde{L}_{23}^{C}-M_{1}^{C33}-M_{3}%
^{C11}+M_{1}^{C32}+M_{1}^{C23}+M_{3}^{C21}+M_{3}^{C12}+R_{132})\nonumber\\
+  &  (1\leftrightarrow3,2\leftrightarrow4).
\end{align}
The integrated w.r.t. $\vec{r}_{5}$ part of the kernel has the form%
\begin{equation}
\mathbf{\mathbf{G}}^{conf}=\mathbf{\mathbf{G}}_{1}^{conf}+\mathbf{\mathbf{G}%
}_{2}^{conf}.
\end{equation}
Here%
\begin{align*}
\mathbf{\mathbf{G}}_{1}^{conf}  &  =(\mathbf{U}_{0^{\dag}3}\mathbf{U}%
_{2^{\dag}1}\mathbf{U}_{4^{\dag}0}-\mathbf{U}_{0^{\dag}12^{\dag}04^{\dag}%
3}-\mathbf{U}_{2^{\dag}34^{\dag}1}(N_{c}^{2}-1))\\
&  \times\frac{1}{4}\left[  \frac{\vec{r}_{14}{}^{2}}{\vec{r}_{01}{}^{2}%
\vec{r}_{04}{}^{2}}\left(  \ln^{2}\left(  \frac{\vec{r}_{02}{}^{2}\vec{r}%
_{14}{}^{2}}{\vec{r}_{04}{}^{2}\vec{r}_{12}{}^{2}}\right)  -\ln^{2}\left(
\frac{\vec{r}_{03}{}^{2}\vec{r}_{14}{}^{2}}{\vec{r}_{04}{}^{2}\vec{r}_{13}%
{}^{2}}\right)  \right)  -\frac{\vec{r}_{13}{}^{2}}{\vec{r}_{01}{}^{2}\vec
{r}_{03}{}^{2}}\ln^{2}\left(  \frac{\vec{r}_{02}{}^{2}\vec{r}_{13}{}^{2}}%
{\vec{r}_{03}{}^{2}\vec{r}_{12}{}^{2}}\right)  \right. \\
&  +\frac{\vec{r}_{23}{}^{2}}{\vec{r}_{03}{}^{2}\vec{r}_{02}{}^{2}}\left(
\ln^{2}\left(  \frac{\vec{r}_{01}{}^{2}\vec{r}_{23}{}^{2}}{\vec{r}_{03}{}%
^{2}\vec{r}_{12}{}^{2}}\right)  -\ln^{2}\left(  \frac{\vec{r}_{03}{}^{2}%
\vec{r}_{24}{}^{2}}{\vec{r}_{04}{}^{2}\vec{r}_{23}{}^{2}}\right)  \right)
-\frac{\vec{r}_{24}{}^{2}}{\vec{r}_{04}{}^{2}\vec{r}_{02}{}^{2}}\ln^{2}\left(
\frac{\vec{r}_{01}{}^{2}\vec{r}_{24}{}^{2}}{\vec{r}_{04}{}^{2}\vec{r}_{12}%
{}^{2}}\right) \\
&  +\left.  \frac{\vec{r}_{34}{}^{2}}{\vec{r}_{04}{}^{2}\vec{r}_{03}{}^{2}%
}\left(  \ln^{2}\left(  \frac{\vec{r}_{01}{}^{2}\vec{r}_{34}{}^{2}}{\vec
{r}_{04}{}^{2}\vec{r}_{13}{}^{2}}\right)  +\ln^{2}\left(  \frac{\vec{r}_{02}%
{}^{2}\vec{r}_{34}{}^{2}}{\vec{r}_{03}{}^{2}\vec{r}_{24}{}^{2}}\right)
\right)  \right]
\end{align*}%
\begin{align}
+  &  (\mathbf{U}_{0^{\dag}1}\mathbf{U}_{2^{\dag}3}\mathbf{U}_{4^{\dag}%
0}-\mathbf{U}_{0^{\dag}14^{\dag}02^{\dag}3}-\mathbf{U}_{2^{\dag}34^{\dag}%
1}(N_{c}^{2}-1))\nonumber\\
\times &  \frac{1}{4}\left[  \frac{\vec{r}_{34}{}^{2}}{\vec{r}_{03}{}^{2}%
\vec{r}_{04}{}^{2}}\left(  \ln^{2}\left(  \frac{\vec{r}_{02}{}^{2}\vec{r}%
_{34}{}^{2}}{\vec{r}_{04}{}^{2}\vec{r}_{23}{}^{2}}\right)  -\ln^{2}\left(
\frac{\vec{r}_{01}{}^{2}\vec{r}_{34}{}^{2}}{\vec{r}_{04}{}^{2}\vec{r}_{13}%
{}^{2}}\right)  \right)  -\frac{\vec{r}_{24}{}^{2}}{\vec{r}_{02}{}^{2}\vec
{r}_{04}{}^{2}}\ln^{2}\left(  \frac{\vec{r}_{03}{}^{2}\vec{r}_{24}{}^{2}}%
{\vec{r}_{04}{}^{2}\vec{r}_{23}{}^{2}}\right)  \right. \nonumber\\
+  &  \frac{\vec{r}_{12}{}^{2}}{\vec{r}_{02}{}^{2}\vec{r}_{01}{}^{2}}\left(
\ln^{2}\left(  \frac{\vec{r}_{01}{}^{2}\vec{r}_{23}{}^{2}}{\vec{r}_{03}{}%
^{2}\vec{r}_{12}{}^{2}}\right)  -\ln^{2}\left(  \frac{\vec{r}_{01}{}^{2}%
\vec{r}_{24}{}^{2}}{\vec{r}_{04}{}^{2}\vec{r}_{12}{}^{2}}\right)  \right)
-\frac{\vec{r}_{13}{}^{2}}{\vec{r}_{03}{}^{2}\vec{r}_{01}{}^{2}}\ln^{2}\left(
\frac{\vec{r}_{01}{}^{2}\vec{r}_{23}{}^{2}}{\vec{r}_{02}{}^{2}\vec{r}_{13}%
{}^{2}}\right) \nonumber\\
+  &  \left.  \frac{\vec{r}_{14}{}^{2}}{\vec{r}_{04}{}^{2}\vec{r}_{01}{}^{2}%
}\left(  \ln^{2}\left(  \frac{\vec{r}_{03}{}^{2}\vec{r}_{14}{}^{2}}{\vec
{r}_{04}{}^{2}\vec{r}_{13}{}^{2}}\right)  +\ln^{2}\left(  \frac{\vec{r}_{01}%
{}^{2}\vec{r}_{24}{}^{2}}{\vec{r}_{02}{}^{2}\vec{r}_{14}{}^{2}}\right)
\right)  \right]  +(1\leftrightarrow3,2\leftrightarrow4).
\end{align}%
\begin{align}
\mathbf{\mathbf{G}}_{2}^{conf}=  &  \frac{N_{c}}{4}(\mathbf{U}_{2^{\dag}%
0}\mathbf{U}_{0^{\dag}34^{\dag}1}-N_{c}\mathbf{U}_{2^{\dag}34^{\dag}1})\left[
\frac{\vec{r}_{13}{}^{2}}{\vec{r}_{01}{}^{2}\vec{r}_{03}{}^{2}}\left(  \ln
^{2}\left(  \frac{\vec{r}_{02}{}^{2}\vec{r}_{13}{}^{2}}{\vec{r}_{03}{}^{2}%
\vec{r}_{12}{}^{2}}\right)  +\ln^{2}\left(  \frac{\vec{r}_{01}{}^{2}\vec
{r}_{23}{}^{2}}{\vec{r}_{02}{}^{2}\vec{r}_{13}{}^{2}}\right)  \right)  \right.
\nonumber\\
-  &  \left.  \frac{\vec{r}_{23}{}^{2}}{\vec{r}_{02}{}^{2}\vec{r}_{03}{}^{2}%
}\ln^{2}\left(  \frac{\vec{r}_{01}{}^{2}\vec{r}_{23}{}^{2}}{\vec{r}_{03}{}%
^{2}\vec{r}_{12}{}^{2}}\right)  -\frac{\vec{r}_{12}{}^{2}}{\vec{r}_{01}{}%
^{2}\vec{r}_{02}{}^{2}}\ln^{2}\left(  \frac{\vec{r}_{01}{}^{2}\vec{r}_{23}%
{}^{2}}{\vec{r}_{03}{}^{2}\vec{r}_{12}{}^{2}}\right)  \right] \nonumber\\
+  &  \frac{N_{c}}{4}(\mathbf{U}_{0^{\dag}3}\mathbf{U}_{2^{\dag}04^{\dag}%
1}-N_{c}\mathbf{U}_{2^{\dag}34^{\dag}1})\left[  \frac{\vec{r}_{24}{}^{2}}%
{\vec{r}_{02}{}^{2}\vec{r}_{04}{}^{2}}\left(  \ln^{2}\left(  \frac{\vec
{r}_{03}{}^{2}\vec{r}_{24}{}^{2}}{\vec{r}_{04}{}^{2}\vec{r}_{23}{}^{2}%
}\right)  +\ln^{2}\left(  \frac{\vec{r}_{02}{}^{2}\vec{r}_{34}{}^{2}}{\vec
{r}_{03}{}^{2}\vec{r}_{24}{}^{2}}\right)  \right)  \right. \nonumber\\
-  &  \left.  \frac{\vec{r}_{23}{}^{2}}{\vec{r}_{02}{}^{2}\vec{r}_{03}{}^{2}%
}\ln^{2}\left(  \frac{\vec{r}_{02}{}^{2}\vec{r}_{34}{}^{2}}{\vec{r}_{04}{}%
^{2}\vec{r}_{23}{}^{2}}\right)  -\frac{\vec{r}_{34}{}^{2}}{\vec{r}_{03}{}%
^{2}\vec{r}_{04}{}^{2}}\ln^{2}\left(  \frac{\vec{r}_{02}{}^{2}\vec{r}_{34}%
{}^{2}}{\vec{r}_{04}{}^{2}\vec{r}_{23}{}^{2}}\right)  \right]
+(1\leftrightarrow3,2\leftrightarrow4).
\end{align}
It was straightforwardly checked that all the integrals of $\mathbf{\mathbf{G}%
}_{s}^{conf}\mathbf{\mathbf{,}}$ $\mathbf{\mathbf{G}}_{a}^{conf},$ and
$\mathbf{\mathbf{G}}^{conf}$ do not have unintegrable singularities.

\subsection{Double dipole}

For symmetric contribution $\mathbf{\mathbf{\tilde{G}}}_{s}^{conf}$ we have%
\begin{equation}
\mathbf{\mathbf{\tilde{G}}}_{s}^{conf}=\mathbf{\mathbf{\tilde{G}}}_{s1}%
^{conf}+n_{f}\mathbf{\tilde{G}}_{q}+\mathbf{\mathbf{\tilde{G}}}_{s2}^{conf},
\end{equation}
where $\mathbf{\tilde{G}}_{q}$ (\ref{Gq}) did not change.%
\begin{align}
\mathbf{\tilde{G}}_{s1}^{conf}=  &  \left(  \left\{  \mathbf{U}_{0^{\dag
}12^{\dag}5}\mathbf{U}_{4^{\dag}35^{\dag}0}-\mathbf{U}_{0^{\dag}5}%
\mathbf{U}_{2^{\dag}15^{\dag}34^{\dag}0}-(5\rightarrow0)\right\}
+(5\leftrightarrow0)\right)  \left(  L_{14}^{C}-L_{13}^{C}+L_{23}^{C}%
-L_{24}^{C}\right) \nonumber\\
+  &  (1\leftrightarrow3,2\leftrightarrow4),\\
\mathbf{\tilde{G}}_{s2}^{conf}=  &  \left(  M_{4}^{C13}-M_{4}^{C31}%
+M_{3}^{C14}-M_{3}^{C41}-M_{4}^{C23}+M_{4}^{C32}-M_{3}^{C24}+M_{3}%
^{C42}\right) \nonumber\\
\times &  \frac{1}{2}(\mathbf{U}_{4^{\dag}0}\mathbf{U}_{0^{\dag}35^{\dag
}12^{\dag}5}+\mathbf{U}_{4^{\dag}0}\mathbf{U}_{0^{\dag}52^{\dag}15^{\dag}%
3}-\mathbf{U}_{0^{\dag}3}\mathbf{U}_{2^{\dag}15^{\dag}04^{\dag}5}%
-\mathbf{U}_{0^{\dag}3}\mathbf{U}_{2^{\dag}54^{\dag}05^{\dag}1}%
+(5\leftrightarrow0))\nonumber\\
+  &  (M_{4}^{C12}-M_{4}^{C21}+M_{3}^{C21}-M_{3}^{C12}-M_{1}^{C34}+M_{1}%
^{C43}-M_{2}^{C43}+M_{2}^{C34})\nonumber\\
\times &  \frac{1}{2}(\mathbf{U}_{0^{\dag}54^{\dag}3}\mathbf{U}_{2^{\dag
}05^{\dag}1}-\mathbf{U}_{0^{\dag}5}\mathbf{U}_{2^{\dag}15^{\dag}34^{\dag}%
0}+(5\leftrightarrow0))+(1\leftrightarrow3,2\leftrightarrow4).
\end{align}
The antisymmetric contribution reads%
\begin{align}
\mathbf{\mathbf{\tilde{G}}}_{a}=  &  \frac{1}{2}(\mathbf{U}_{5^{\dag}%
3}\mathbf{U}_{0^{\dag}12^{\dag}04^{\dag}5}+\mathbf{U}_{5^{\dag}3}%
\mathbf{U}_{0^{\dag}54^{\dag}02^{\dag}1}-\mathbf{U}_{4^{\dag}5}\mathbf{U}%
_{0^{\dag}12^{\dag}05^{\dag}3}-\mathbf{U}_{4^{\dag}5}\mathbf{U}_{0^{\dag
}35^{\dag}02^{\dag}1}-(5\leftrightarrow0))\nonumber\\
\times &  (M_{4}^{23}+M_{3}^{24}-M_{4}^{13}-M_{3}^{14}-M_{4}^{31}+M_{4}%
^{32}-M_{3}^{41}+M_{3}^{42}-R_{341}+R_{342})\nonumber\\
+  &  \frac{1}{2}(\mathbf{U}_{0^{\dag}54^{\dag}3}\mathbf{U}_{2^{\dag}05^{\dag
}1}-\mathbf{U}_{0^{\dag}52^{\dag}1}\mathbf{U}_{4^{\dag}05^{\dag}3}%
+\mathbf{U}_{5^{\dag}0}\mathbf{U}_{0^{\dag}12^{\dag}54^{\dag}3}-\mathbf{U}%
_{0^{\dag}5}\mathbf{U}_{2^{\dag}15^{\dag}34^{\dag}0}-(5\leftrightarrow
0))\nonumber\\
\times &  (M_{4}^{C11}-M_{3}^{C11}+M_{3}^{C12}-M_{4}^{C12}+M_{3}^{C21}%
-M_{4}^{C21}-M_{3}^{C22}+M_{4}^{C22})+(1\leftrightarrow3,2\leftrightarrow4).
\end{align}
The contribution with 1 gluon crossing the shockwave has the form
\begin{align}
\mathbf{\tilde{G}}^{conf}=  &  \frac{1}{4}(\mathbf{U}_{2^{\dag}0}%
\mathbf{U}_{0^{\dag}14^{\dag}3}+\mathbf{U}_{2^{\dag}0}\mathbf{U}_{0^{\dag
}34^{\dag}1}-\mathbf{U}_{0^{\dag}1}\mathbf{U}_{2^{\dag}04^{\dag}3}%
-\mathbf{U}_{0^{\dag}1}\mathbf{U}_{2^{\dag}34^{\dag}0})\nonumber\\
\times &  \left[  \frac{\vec{r}_{12}{}^{2}}{\vec{r}_{01}{}^{2}\vec{r}_{02}%
{}^{2}}\left(  \ln^{2}\left(  \frac{\vec{r}_{02}{}^{2}\vec{r}_{14}{}^{2}}%
{\vec{r}_{04}{}^{2}\vec{r}_{12}{}^{2}}\right)  -\ln^{2}\left(  \frac{\vec
{r}_{02}{}^{2}\vec{r}_{13}{}^{2}}{\vec{r}_{03}{}^{2}\vec{r}_{12}{}^{2}%
}\right)  -\ln^{2}\left(  \frac{\vec{r}_{01}{}^{2}\vec{r}_{23}{}^{2}}{\vec
{r}_{03}{}^{2}\vec{r}_{12}{}^{2}}\right)  +\ln^{2}\left(  \frac{\vec{r}_{01}%
{}^{2}\vec{r}_{24}{}^{2}}{\vec{r}_{04}{}^{2}\vec{r}_{12}{}^{2}}\right)
\right)  \right. \nonumber\\
+  &  \left.  \left(  \frac{\vec{r}_{13}{}^{2}}{\vec{r}_{01}{}^{2}\vec{r}%
_{03}{}^{2}}+\frac{\vec{r}_{23}{}^{2}}{\vec{r}_{02}{}^{2}\vec{r}_{03}{}^{2}%
}\right)  \ln^{2}\left(  \frac{\vec{r}_{01}{}^{2}\vec{r}_{23}{}^{2}}{\vec
{r}_{02}{}^{2}\vec{r}_{13}{}^{2}}\right)  -\left(  \frac{\vec{r}_{14}{}^{2}%
}{\vec{r}_{01}{}^{2}\vec{r}_{04}{}^{2}}+\frac{\vec{r}_{24}{}^{2}}{\vec{r}%
_{02}{}^{2}\vec{r}_{04}{}^{2}}\right)  \ln^{2}\left(  \frac{\vec{r}_{01}{}%
^{2}\vec{r}_{24}{}^{2}}{\vec{r}_{02}{}^{2}\vec{r}_{14}{}^{2}}\right)  \right]
\nonumber\\
+  &  \frac{1}{4}\left(  N_{c}\mathbf{U}_{0^{\dag}12^{\dag}04^{\dag}3}%
+N_{c}\mathbf{U}_{0^{\dag}34^{\dag}02^{\dag}1}-2\mathbf{U}_{2^{\dag}%
1}\mathbf{U}_{34^{\dag}}\right)  \left[  \frac{\vec{r}_{24}{}^{2}}{\vec
{r}_{02}{}^{2}\vec{r}_{04}{}^{2}}\ln^{2}\left(  \frac{\vec{r}_{02}{}^{2}%
\vec{r}_{34}{}^{2}}{\vec{r}_{03}{}^{2}\vec{r}_{24}{}^{2}}\right)  \right.
\nonumber\\
+  &  \left.  \frac{\vec{r}_{13}{}^{2}}{\vec{r}_{01}{}^{2}\vec{r}_{03}{}^{2}%
}\ln^{2}\left(  \frac{\vec{r}_{02}{}^{2}\vec{r}_{13}{}^{2}}{\vec{r}_{03}{}%
^{2}\vec{r}_{12}{}^{2}}\right)  -\frac{\vec{r}_{14}{}^{2}}{\vec{r}_{01}{}%
^{2}\vec{r}_{04}{}^{2}}\left(  \ln^{2}\left(  \frac{\vec{r}_{02}{}^{2}\vec
{r}_{14}{}^{2}}{\vec{r}_{04}{}^{2}\vec{r}_{12}{}^{2}}\right)  +\ln^{2}\left(
\frac{\vec{r}_{01}{}^{2}\vec{r}_{34}{}^{2}}{\vec{r}_{03}{}^{2}\vec{r}_{14}%
{}^{2}}\right)  \right)  \right] \nonumber\\
+  &  (1\leftrightarrow3,2\leftrightarrow4).
\end{align}
As for the quadrupole, it was straightforwardly checked that all the integrals
of $\mathbf{\mathbf{\tilde{G}}}_{s}^{conf}\mathbf{\mathbf{,}}$
$\mathbf{\mathbf{\tilde{G}}}_{a}^{conf},$ and $\mathbf{\mathbf{\tilde{G}}%
}^{conf}$ do not have unintegrable singularities.

\section{Checks}

There are two checks which can be done for the results of this paper. The
evolution equations for the quadrupole and double dipole operators can be
obtained from the NLO JIMWLK hamiltonian \cite{Kovner:2014lca} and the general
evolution equations from \cite{sch}.

In this paper the following two checks were done. First, quadrupole kernels
(\ref{Gdefinition}) and (\ref{GdefinitionC}) go into BK ones (\ref{NLO_BK})
and (\ref{NLOBKC}) in the dipole limits $1\rightarrow2,\,2\rightarrow
3,\,3\rightarrow4,$ and $4\rightarrow1$. Double dipole kernels
(\ref{Gtildedefinition}) and (\ref{GtildedefinitionC}) also have the correct
dipole limits $1\rightarrow2\,$and $3\rightarrow4.$ In these limits they also
go into the BK ones (\ref{NLO_BK}) and (\ref{NLOBKC}) times $N_{c}$. This
statement can be checked straightforwardly going to the dipole limits in
explicit expressions (\ref{Gdefinition}), (\ref{GdefinitionC}),
(\ref{Gtildedefinition}), and (\ref{GtildedefinitionC}). Our kernels match the
Balitsky-Fadin-Kuraev-Lipatov NLO kernel \cite{Fadin:2009gh} in these limits.

The second check is that in $SU(3)$ our kernels respect the identity%
\begin{equation}
B_{123}\equiv\mathbf{U}_{12^{\dag}}\mathbf{U}_{34^{\dag}}-\mathbf{U}%
_{12^{\dag}34^{\dag}}, \label{3qwl=dd-q}%
\end{equation}
where $B_{123}\ $is the 3-quark Wilson loop (baryon) operator defined as
\begin{equation}
B_{123}\equiv U_{1}\cdot U_{2}\cdot U_{3}\equiv\varepsilon^{i^{\prime
}j^{\prime}h^{\prime}}\varepsilon_{ijh}U_{1i^{\prime}}^{i}U_{2j^{\prime}}%
^{j}U_{3h^{\prime}}^{h}. \label{B}%
\end{equation}
The evolution equations for the l.h.s. of (\ref{3qwl=dd-q}) in the standard
and quasi-conformal forms are given in \cite{bg}. They read%
\begin{align*}
\frac{\partial B_{123}}{\partial\eta}  &  =\frac{\alpha_{s}(\mu^{2})}{8\pi
^{2}}\int d\vec{r}_{0}\left[  (B_{100}B_{320}+B_{200}B_{310}-B_{300}%
B_{210}-6B_{123})\frac{{}}{{}}\right. \\
&  \times\left\{  \frac{\vec{r}_{12}^{\,\,2}}{\vec{r}_{01}^{\,\,2}\vec{r}%
_{02}^{\,\,2}}-{\frac{3\alpha_{s}}{4\pi}}\!\beta\left[  \ln\left(  \frac
{\vec{r}_{01}^{\,\,2}}{\vec{r}_{02}^{\,\,2}}\right)  \left(  \frac{1}{\vec
{r}_{02}^{\,\,2}}-\frac{1}{\vec{r}_{01}^{\,\,2}}\right)  -\frac{\vec{r}%
_{12}^{\,\,2}}{\vec{r}_{01}^{\,\,2}\vec{r}_{02}^{\,\,2}}\ln\left(  \frac
{\vec{r}_{12}^{\,\,2}}{\tilde{\mu}^{2}}\right)  \right]  \right\} \\
&  -{\frac{\alpha_{s}}{\pi}}\!~\ln\frac{\vec{r}_{20}^{\,\,2}}{\vec{r}%
_{21}^{\,\,2}}\ln\frac{\vec{r}_{10}^{\,\,2}}{\vec{r}_{21}^{\,\,2}}\left\{
\frac{1}{2}\left[  \frac{\vec{r}_{13}^{\,\,2}}{\vec{r}_{10}^{\,\,2}\vec
{r}_{30}^{\,\,\,2}}-\frac{\vec{r}_{32}^{\,2}}{\vec{r}_{30}^{\,2}\vec{r}%
_{20}^{\,\,\,2}}\right]  \left(  B_{100}B_{320}-B_{200}B_{310}\right)  \right.
\\
&  -\left.  \left.  \frac{\vec{r}_{12}^{\,2}}{\vec{r}_{10}^{\,2}\vec{r}%
_{20}^{\,\,\,2}}\left(  9B_{123}-\frac{1}{2}\left[  2\left(  B_{100}%
B_{320}+B_{200}B_{130}\right)  -B_{300}B_{120}\right]  \right)  \right\}
+(1\leftrightarrow3)+(2\leftrightarrow3)\right]
\end{align*}%
\begin{align}
&  -{\frac{\alpha_{s}^{2}n_{f}}{16\pi^{4}}}\!\int\!d\vec{r}_{0}d\vec{r}%
_{5}\left[  \left\{  \left(  \frac{1}{3}(U_{1}U_{0}{}^{\dag}U_{5}+U_{5}U_{0}%
{}^{\dag}U_{1})\cdot U_{2}\cdot U_{3}-\frac{1}{9}B_{123}tr(U_{0}{}^{\dag}%
U_{5})\right.  \right.  \right. \nonumber\\
&  +(U_{1}U_{0}{}^{\dag}U_{2})\cdot U_{3}\cdot U_{5}+\frac{1}{6}B_{123}%
-\frac{1}{4}(B_{013}B_{002}+B_{001}B_{023}-B_{012}B_{003})\nonumber\\
&  +\left.  \left.  \left.  \frac{{}}{{}}(1\leftrightarrow2)\right)
+(0\leftrightarrow5)\right\}  L_{12}^{q}+(1\leftrightarrow3)+(2\leftrightarrow
3)\right] \nonumber\\
&  -{\frac{\alpha_{s}^{2}}{8\pi^{4}}}\!\int\!d\vec{r}_{0}d\vec{r}_{5}~\left[
\mathbf{\{}\tilde{L}_{12}\left(  U_{0}U_{5}{}^{\dag}U_{2}\right)  \cdot\left(
U_{1}U_{0}{}^{\dag}U_{5}\right)  \cdot U_{3}\frac{{}}{{}}\right. \nonumber\\
&  \mathbf{+}L_{12}\left[  \left(  U_{0}U_{5}{}^{\dag}U_{2}\right)
\cdot\left(  U_{1}U_{0}{}^{\dag}U_{5}\right)  \cdot U_{3}+tr\left(  U_{0}%
U_{5}{}^{\dag}\right)  \left(  U_{1}U_{0}{}^{\dag}U_{2}\right)  \cdot
U_{3}\cdot U_{5}\frac{{}}{{}}\right. \nonumber\\
&  -\left.  \frac{3}{4}[B_{155}B_{235}+B_{255}B_{135}-B_{355}B_{125}]+\frac
{1}{2}B_{123}\right] \nonumber\\
&  +(M_{13}-M_{12}-M_{23}+M_{2}^{13})\left[  (U_{0}U_{5}{}^{\dag}U_{3}%
)\cdot(U_{2}U_{0}{}^{\dag}U_{1})\cdot U_{5}+(U_{1}U_{0}{}^{\dag}U_{2}%
)\cdot(U_{3}U_{5}{}^{\dag}U_{0})\cdot U_{5}\right] \nonumber\\
&  +\left.  (\text{all 5 permutations}\,1\leftrightarrow2\leftrightarrow
3)\}+\frac{{}}{{}}(0\leftrightarrow5)\right]  , \label{3qwleeq}%
\end{align}
where $L,\,L^{q},$ $M_{i}^{jk}$ and $M_{ij}$ are introduced in (\ref{L12}),
(\ref{L12q}), (\ref{M}), and (\ref{Mij}).(\ref{3qwleeq}) and (\ref{3qwleeqC}%
).
\begin{align*}
\frac{\partial B_{123}^{conf}}{\partial\eta}  &  =\frac{\alpha_{s}\left(
\mu^{2}\right)  }{8\pi^{2}}\int d\vec{r}_{0}\left[  ((B_{100}B_{320}%
+B_{200}B_{310}-B_{300}B_{210})-6B_{123})^{conf}\frac{{}}{{}}\right. \\
\times &  \left.  \left(  \frac{\vec{r}_{12}^{\,\,2}}{\vec{r}_{01}^{\,\,2}%
\vec{r}_{02}^{\,\,2}}-\!\frac{3\alpha_{s}}{4\pi}\beta\left[  \ln\left(
\frac{\vec{r}_{01}^{\,\,2}}{\vec{r}_{02}^{\,\,2}}\right)  \left(  \frac
{1}{\vec{r}_{02}^{\,\,2}}-\frac{1}{\vec{r}_{01}^{\,\,2}}\right)  -\frac
{\vec{r}_{12}^{\,\,2}}{\vec{r}_{01}^{\,\,2}\vec{r}_{02}^{\,\,2}}\ln\left(
\frac{\vec{r}_{12}^{\,\,2}}{\tilde{\mu}^{2}}\right)  \right]  \right)
+(1\leftrightarrow3)+(2\leftrightarrow3)\right] \\
-  &  {\frac{\alpha_{s}^{2}}{32\pi^{3}}}\!\int\!d\vec{r}_{0}\left(
B_{003}B_{012}\left[  \frac{\vec{r}_{32}{}^{2}}{\vec{r}_{03}{}^{2}\vec{r}%
_{02}{}^{2}}\ln^{2}\left(  \frac{\vec{r}_{32}{}^{2}\vec{r}_{10}{}^{2}}{\vec
{r}_{13}{}^{2}\vec{r}_{20}{}^{2}}\right)  -\frac{\vec{r}_{12}{}^{2}}{\vec
{r}_{01}{}^{2}\vec{r}_{02}{}^{2}}\ln^{2}\left(  \frac{\vec{r}_{12}{}^{2}%
\vec{r}_{30}{}^{2}}{\vec{r}_{13}{}^{2}\vec{r}_{20}{}^{2}}\right)  \right]
\right. \\
+  &  \left.  \frac{{}}{{}}(\text{all 5 permutations}\,1\leftrightarrow
2\leftrightarrow3)\right) \\
-  &  {\frac{\alpha_{s}^{2}n_{f}}{16\pi^{4}}}\!\int\!d\vec{r}_{0}d\vec{r}%
_{5}\left[  \left\{  \left(  \frac{1}{3}(U_{1}U_{0}{}^{\dag}U_{5}+U_{5}U_{0}%
{}^{\dag}U_{1})\cdot U_{2}\cdot U_{3}-\frac{1}{9}B_{123}tr(U_{0}{}^{\dag}%
U_{5})\right.  \right.  \right. \\
+  &  (U_{1}U_{0}{}^{\dag}U_{2})\cdot U_{3}\cdot U_{5}+\frac{1}{6}%
B_{123}-\frac{1}{4}(B_{013}B_{002}+B_{001}B_{023}-B_{012}B_{003})\\
+  &  \left.  \left.  \left.  \frac{{}}{{}}(1\leftrightarrow2)\right)
+(0\leftrightarrow5)\right\}  L_{12}^{q}+(1\leftrightarrow3)+(2\leftrightarrow
3)\right]
\end{align*}%
\begin{align}
-  &  {\frac{\alpha_{s}^{2}}{8\pi^{4}}}\!\int\!d\vec{r}_{0}d\vec{r}%
_{5}~\left(  \left\{  \tilde{L}_{12}^{C}\left(  U_{0}U_{5}{}^{\dag}%
U_{2}\right)  \cdot\left(  U_{1}U_{0}{}^{\dag}U_{5}\right)  \cdot U_{3}%
\frac{{}}{{}}\right.  \right. \nonumber\\
\mathbf{+}  &  L_{12}^{C}\left[  \left(  U_{0}U_{5}{}^{\dag}U_{2}\right)
\cdot\left(  U_{1}U_{0}{}^{\dag}U_{5}\right)  \cdot U_{3}+tr\left(  U_{0}%
U_{5}{}^{\dag}\right)  \left(  U_{1}U_{0}{}^{\dag}U_{2}\right)  \cdot
U_{3}\cdot U_{5}\frac{{}}{{}}\right. \nonumber\\
-  &  \left.  \frac{3}{4}[B_{155}B_{235}+B_{255}B_{135}-B_{355}B_{125}%
]+\frac{1}{2}B_{123}\right] \nonumber\\
+  &  \tilde{M}_{12}^{C}\left[  \left(  U_{0}U_{5}{}^{\dag}U_{3}\right)
\cdot\left(  U_{2}U_{0}{}^{\dag}U_{1}\right)  \cdot U_{5}+\left(  U_{1}U_{0}%
{}^{\dag}U_{2}\right)  \cdot\left(  U_{3}U_{5}{}^{\dag}U_{0}\right)  \cdot
U_{5}\right] \nonumber\\
+  &  \left.  \left.  \frac{{}}{{}}(\text{all 5 permutations}%
\,1\leftrightarrow2\leftrightarrow3)\right\}  +(0\leftrightarrow5)\right)  ,
\label{3qwleeqC}%
\end{align}
where $L^{C}$ is defined in (\ref{LC}), $\tilde{L}^{C}$ --- in (\ref{LCtilde}%
), and
\begin{align}
M_{12}^{C}  &  =\frac{\vec{r}_{12}{}^{2}}{16\vec{r}_{02}{}^{2}\vec{r}_{05}%
{}^{2}\vec{r}_{15}{}^{2}}\ln\left(  \frac{\vec{r}_{01}{}^{2}\vec{r}_{02}{}%
^{2}\vec{r}_{35}{}^{4}{}}{\vec{r}_{03}{}^{4}{}\vec{r}_{15}{}^{2}\vec{r}_{25}%
{}^{2}}\right)  +\frac{\vec{r}_{12}{}^{2}}{16\vec{r}_{01}{}^{2}\vec{r}_{05}%
{}^{2}\vec{r}_{25}{}^{2}}\ln\left(  \frac{\vec{r}_{03}{}^{4}{}\vec{r}_{05}%
{}^{4}{}\vec{r}_{12}{}^{4}{}\vec{r}_{25}{}^{2}}{\vec{r}_{01}{}^{2}\vec{r}%
_{02}{}^{6}{}\vec{r}_{15}{}^{2}\vec{r}_{35}{}^{4}{}}\right) \nonumber\\
&  +\frac{\vec{r}_{23}{}^{2}}{16\vec{r}_{02}{}^{2}\vec{r}_{05}{}^{2}\vec
{r}_{35}{}^{2}}\ln\left(  \frac{\vec{r}_{01}{}^{4}{}\vec{r}_{03}{}^{2}\vec
{r}_{25}{}^{6}{}\vec{r}_{35}{}^{2}}{\vec{r}_{02}{}^{2}\vec{r}_{05}{}^{4}{}%
\vec{r}_{15}{}^{4}{}\vec{r}_{23}{}^{4}{}}\right)  +\frac{\vec{r}_{23}{}^{2}%
}{16\vec{r}_{03}{}^{2}\vec{r}_{05}{}^{2}\vec{r}_{25}{}^{2}}\ln\left(
\frac{\vec{r}_{02}{}^{2}\vec{r}_{03}{}^{2}\vec{r}_{15}{}^{4}{}}{\vec{r}_{01}%
{}^{4}{}\vec{r}_{25}{}^{2}\vec{r}_{35}{}^{2}}\right) \nonumber\\
&  +\frac{\vec{r}_{13}{}^{2}}{16\vec{r}_{03}{}^{2}\vec{r}_{05}{}^{2}\vec
{r}_{15}{}^{2}}\ln\left(  \frac{\vec{r}_{02}{}^{4}{}\vec{r}_{15}{}^{2}\vec
{r}_{35}{}^{2}}{\vec{r}_{01}{}^{2}\vec{r}_{03}{}^{2}\vec{r}_{25}{}^{4}{}%
}\right)  +\frac{\vec{r}_{13}{}^{2}}{16\vec{r}_{01}{}^{2}\vec{r}_{05}{}%
^{2}\vec{r}_{35}{}^{2}}\ln\left(  \frac{\vec{r}_{02}{}^{4}{}\vec{r}_{15}{}%
^{2}\vec{r}_{35}{}^{2}}{\vec{r}_{01}{}^{2}\vec{r}_{03}{}^{2}\vec{r}_{25}{}%
^{4}{}}\right) \nonumber\\
&  +\frac{\vec{r}_{03}{}^{2}\vec{r}_{12}{}^{2}}{8\vec{r}_{01}{}^{2}\vec
{r}_{02}{}^{2}\vec{r}_{05}{}^{2}\vec{r}_{35}{}^{2}}\ln\left(  \frac{\vec
{r}_{01}{}^{2}\vec{r}_{03}{}^{2}\vec{r}_{25}{}^{4}{}}{\vec{r}_{02}{}^{2}%
\vec{r}_{05}{}^{2}\vec{r}_{12}{}^{2}\vec{r}_{35}{}^{2}}\right)  +\frac{\vec
{r}_{23}{}^{2}\vec{r}_{12}{}^{2}}{8\vec{r}_{01}{}^{2}\vec{r}_{02}{}^{2}\vec
{r}_{25}{}^{2}\vec{r}_{35}{}^{2}}\ln\left(  \frac{\vec{r}_{02}{}^{2}\vec
{r}_{12}{}^{2}\vec{r}_{35}{}^{2}}{\vec{r}_{01}{}^{2}\vec{r}_{23}{}^{2}\vec
{r}_{25}{}^{2}}\right) \nonumber\\
&  +\frac{\vec{r}_{15}{}^{2}\vec{r}_{23}{}^{2}}{8\vec{r}_{01}{}^{2}\vec
{r}_{05}{}^{2}\vec{r}_{25}{}^{2}\vec{r}_{35}{}^{2}}\ln\left(  \frac{\vec
{r}_{01}{}^{2}\vec{r}_{05}{}^{2}\vec{r}_{23}{}^{2}\vec{r}_{25}{}^{2}}{\vec
{r}_{02}{}^{4}{}\vec{r}_{15}{}^{2}\vec{r}_{35}{}^{2}}\right)  . \label{Mc}%
\end{align}

In order to check identity (\ref{3qwl=dd-q}) one has to rewrite the evolution
of its l.h.s. in the same operator basis as the r.h.s. To this end one can use
the $SU(3)$ identities%
\begin{align*}
0=  &  [U_{0}\cdot U_{1}\cdot U_{2}tr\left(  U_{0}{}^{\dag}U_{5}\right)
tr\left(  U_{5}{}^{\dag}U_{3}\right) \\
-  &  tr\left(  U_{5}U_{0}{}^{\dag}\right)  \left(  U_{1}U_{5}{}^{\dag}%
U_{3}+U_{3}U_{5}{}^{\dag}U_{1}\right)  \cdot U_{0}\cdot U_{2}+\left(
U_{0}U_{5}{}^{\dag}U_{1}\right)  \cdot\left(  U_{3}U_{0}{}^{\dag}U_{5}\right)
\cdot U_{2}\\
+  &  \left(  U_{1}U_{5}{}^{\dag}U_{0}\right)  \cdot\left(  U_{5}U_{0}{}%
^{\dag}U_{3}\right)  \cdot U_{2}+(1\leftrightarrow2)]-(5\leftrightarrow0),\\
0=  &  2tr\left(  U_{5}U_{0}{}^{\dag}\right)  \left(  U_{2}U_{5}{}^{\dag}%
U_{3}+U_{3}U_{5}{}^{\dag}U_{2}\right)  \cdot U_{0}\cdot U_{1}\\
+  &  \left(  U_{0}U_{5}{}^{\dag}U_{1}+U_{1}U_{5}{}^{\dag}U_{0}\right)
\cdot\left(  U_{2}U_{0}{}^{\dag}U_{3}+U_{3}U_{0}{}^{\dag}U_{2}\right)  \cdot
U_{5}\\
+  &  \left(  U_{0}U_{5}{}^{\dag}U_{2}-U_{2}U_{5}{}^{\dag}U_{0}\right)
\cdot\left(  U_{3}U_{0}{}^{\dag}U_{1}-U_{1}U_{0}{}^{\dag}U_{3}\right)  \cdot
U_{5}\\
+  &  \left(  U_{0}U_{5}{}^{\dag}U_{3}-U_{3}U_{5}{}^{\dag}U_{0}\right)
\cdot\left(  U_{2}U_{0}{}^{\dag}U_{1}-U_{1}U_{0}{}^{\dag}U_{2}\right)  \cdot
U_{5}-(5\leftrightarrow0)
\end{align*}
in the antisymmetric color structures and then
\begin{equation}
U_{i}\cdot U_{j}\cdot U_{k}=(U_{i}U_{l}^{\dag})\cdot(U_{j}U_{l}^{\dag}%
)\cdot(U_{k}U_{l}^{\dag})=(U_{l}^{\dag}U_{i})\cdot(U_{l}^{\dag}U_{j}%
)\cdot(U_{l}^{\dag}U_{k}) \label{3qWlidentity}%
\end{equation}
with $l=2$ to express $U_{2}$ in terms of $U_{2}^{\dag}.$ After that one can
expand the products of Levi-Civita symbols as%
\begin{equation}
\varepsilon_{ijh}\varepsilon^{i^{\prime}j^{\prime}h^{\prime}}=\left\vert
\begin{tabular}
[c]{lll}%
$\delta_{i}^{i^{\prime}}$ & $\delta_{i}^{j^{\prime}}$ & $\delta_{i}%
^{h^{\prime}}$\\
$\delta_{j}^{i^{\prime}}$ & $\delta_{j}^{j^{\prime}}$ & $\delta_{j}%
^{h^{\prime}}$\\
$\delta_{h}^{i^{\prime}}$ & $\delta_{h}^{j^{\prime}}$ & $\delta_{h}%
^{h^{\prime}}$%
\end{tabular}
\right\vert \label{ee}%
\end{equation}
and see that (\ref{3qwl=dd-q}) is satisfied.

\section{Discussion and conclusion}

This paper presents the evolution equations for the double dipole and
quadrupole operators in the standard (\ref{Gdefinition}),
(\ref{Gtildedefinition}) and quasi-conformal forms (\ref{GdefinitionC}),
(\ref{GtildedefinitionC}). They have correct dipole limits and in SU(3) obey
group identity (\ref{3qwl=dd-q}) with the corresponding evolution equations
for the 3QWL operator obtained in \cite{bg}. This fact ensures the correctness
of all the 3 results. To construct the composite operators, prescription
(\ref{prescription}) was used. It was proposed in \cite{Balitsky:2009xg}\ for
the dipole and proved in \cite{NLOJIMWLKonformal}. Here it gave the
quasi-conformal kernels for both double dipole and quadrupole operators, thus
being checked by the specific calculation of the evolution of the 4-point operators.

Unlike the dipole and the 3QWL operators, the evolution of the quadrupole and
the double dipole ones generates several operators in the virtual part.
Indeed, the virtual gluons do not change the color structure of a dipole or a
baryon. New color structures appear in the evolution of these operators only
when the gluons cross the shockwave. Therefore, one can write the virtual part
of the evolution equations for them without calculation. The double dipole and
the quadrupole, on the contrary, mix in the virtual part with each other and
with the double dipoles and quadrupoles with the other order of the Wilson
lines. Therefore they had to be calculated explicitly. Using the evolution
equations for Wilson lines from \cite{Balitsky:2013fea}\ in this calculation,
one encounters ill-defined integrals which were treated here so as to obtain
the known result for the dipole and the 3QWL operators. Although such
treatment gave the equations satisfying all the checks, it is important to
have the initial expressions with the regularization of the IR singularities
and to check the results of this paper. Such checks may be performed starting
from the evolution equations found in \cite{Kovner:2014lca} and \cite{sch}.

The equations obtained in this paper may be used to derive the NLO evolution
equation for Weizs\"{a}cker-Williams\ gluon distribution. This work is in
progress. They can also be important in the analysis of higher than dipole
Fock components of the virtual photon in the diffractive processes.

\acknowledgments I would like to thank I. Balitsky for proposing this work. I
am also grateful to I. Balitsky, M. G. Kozlov, R. N. Lee, A. I. Milstein, and
A.V. Reznichenko for helpful discussions and to the Dynasty foundation for
financial support. The study was also supported by the Russian Fund for Basic
Research grant 13-02-01023 and president scholarship 171.2015.2.

\end{document}